\documentclass[reprint,
superscriptaddress,
nobibnotes,
amsmath,amssymb,
aps,
]{revtex4-2}

\usepackage[utf8]{inputenc} % allow utf-8 input
\usepackage[T1]{fontenc}    % use 8-bit T1 fonts
\usepackage{hyperref}       % hyperlinks
\usepackage{url}            % simple URL typesetting
\usepackage{amsfonts}       % blackboard math symbols
\usepackage{amsmath,amsthm,amssymb} % math packages
\usepackage{bm}% bold math
\usepackage{bbm}% bold math
\usepackage{dcolumn}% Align table columns on decimal point
\usepackage{nicefrac}       % compact symbols for 1/2, etc.
\usepackage{graphicx}
\usepackage{xcolor, etoolbox}
\usepackage{mathtools}
\usepackage{mathrsfs}
\usepackage{ragged2e}
\usepackage{microtype}      % microtypography
\usepackage{braket}      % Dirac Bra-Ket notation
\usepackage{enumitem}
\usepackage{xspace}
\usepackage{dsfont}
\usepackage{float}

\DeclarePairedDelimiter\abs{\lvert}{\rvert}
\renewcommand{\vec}[1]{\boldsymbol{#1}}

\hypersetup{
    colorlinks,
    linkcolor=blue,
    citecolor=blue,
    urlcolor=blue,
    breaklinks=true 
}

\begin{document}
\title{Curved and expanding spacetime geometries in Bose-Einstein condensates}

\author{Mireia Tolosa-Sime\'on}
    \email{mtolosa@thp.uni-koeln.de}
    \affiliation{Institut f\"{u}r Theoretische Physik, Universit\"{a}t Heidelberg, \\ Philosophenweg 16, 69120 Heidelberg, Germany}
\author{\'Alvaro Parra-L\'opez}
    \email{alvaparr@ucm.es}
    \affiliation{Institut f\"{u}r Theoretische Physik, Universit\"{a}t Heidelberg, \\ Philosophenweg 16, 69120 Heidelberg, Germany}
    \affiliation{Departamento de F\'isica Te\'orica and IPARCOS, Facultad de Ciencias F\'isicas, Universidad Complutense de Madrid, Ciudad Universitaria, 28040 Madrid, Spain}
\author{Natalia S\'anchez-Kuntz}
    \email{sanchez@thphys.uni-heidelberg.de}
    \affiliation{Institut f\"{u}r Theoretische Physik, Universit\"{a}t Heidelberg, \\ Philosophenweg 16, 69120 Heidelberg, Germany}
\author{Tobias Haas}
    \email{t.haas@thphys.uni-heidelberg.de}
    \affiliation{Institut f\"{u}r Theoretische Physik, Universit\"{a}t Heidelberg, \\ Philosophenweg 16, 69120 Heidelberg, Germany}
\author{Celia Viermann}
    \email{curvedspacetime@matterwave.de}
    \affiliation{Kirchhoff-Institut f\"{u}r Physik, Universit\"{a}t Heidelberg, \\
    Im Neuenheimer Feld 227, 69120 Heidelberg, Germany}
\author{Marius Sparn}
    \email{curvedspacetime@matterwave.de}
    \affiliation{Kirchhoff-Institut f\"{u}r Physik, Universit\"{a}t Heidelberg, \\
    Im Neuenheimer Feld 227, 69120 Heidelberg, Germany}
\author{Nikolas Liebster}
    \email{curvedspacetime@matterwave.de}
    \affiliation{Kirchhoff-Institut f\"{u}r Physik, Universit\"{a}t Heidelberg, \\
    Im Neuenheimer Feld 227, 69120 Heidelberg, Germany}  
\author{Maurus Hans}
    \email{curvedspacetime@matterwave.de}
    \affiliation{Kirchhoff-Institut f\"{u}r Physik, Universit\"{a}t Heidelberg, \\
    Im Neuenheimer Feld 227, 69120 Heidelberg, Germany}
\author{Elinor Kath}
    \email{curvedspacetime@matterwave.de}
    \affiliation{Kirchhoff-Institut f\"{u}r Physik, Universit\"{a}t Heidelberg, \\
    Im Neuenheimer Feld 227, 69120 Heidelberg, Germany}  
\author{Helmut Strobel}
    \email{curvedspacetime@matterwave.de}
    \affiliation{Kirchhoff-Institut f\"{u}r Physik, Universit\"{a}t Heidelberg, \\
    Im Neuenheimer Feld 227, 69120 Heidelberg, Germany}    
\author{Markus K. Oberthaler}
    \email{curvedspacetime@matterwave.de}
    \affiliation{Kirchhoff-Institut f\"{u}r Physik, Universit\"{a}t Heidelberg, \\
    Im Neuenheimer Feld 227, 69120 Heidelberg, Germany}    
\author{Stefan Floerchinger}
    \email{stefan.floerchinger@uni-jena.de}
    \affiliation{Institut f\"{u}r Theoretische Physik, Universit\"{a}t Heidelberg, \\ Philosophenweg 16, 69120 Heidelberg, Germany}
    \affiliation{Theoretisch-Physikalisches Institut, Friedrich-Schiller-Universit\"{a}t Jena,\\
    Max-Wien-Platz 1, 07743 Jena, Germany}
    
\date{\today}

\begin{abstract}
Phonons have the characteristic linear dispersion relation of massless relativistic particles. They arise as low energy excitations of Bose-Einstein condensates and, in nonhomogeneous situations, are governed by a space- and time-dependent acoustic metric. We discuss how this metric can be experimentally designed to realize curved spacetime geometries, in particular, expanding Friedmann-Lemaître-Robertson-Walker cosmologies, with negative, vanishing, or positive spatial curvature. A nonvanishing Hubble rate can be obtained through a time-dependent scattering length of the background condensate. For relativistic quantum fields this leads to the phenomenon of particle production, which we describe in detail. We explain how particle production and other interesting features of quantum field theory in curved spacetime can be tested in terms of experimentally accessible correlation functions.
\end{abstract}

\maketitle

\section{Introduction}
\label{sec:Introduction}
When studying quantum field theory in curved spacetime, many interesting phenomena and peculiarities arise. Most prominently, allowing for a time-dependence in the metric, which is naturally the case in cosmology, leads to the phenomenon of particle production \cite{Parker1969,Birrell1982,Mukhanov2007,Weinberg2008}. Although indirect signatures of this effect can be observed, for example in the cosmic microwave background, any direct detection is an open challenge.

In recent years, an analogy of this phenomenon has been studied in the context of Bose-Einstein condensates (BECs), as an integral part of the analog gravity program (see \cite{Barcelo2011,Visser2002,Novello2002} for introductions). More precisely, the linear phononic excitations on top of the ground state of a BEC obey a Klein-Gordon equation for a scalar quantum field in curved spacetime in the acoustic approximation (that is neglecting quantum pressure) \cite{Volovik2009}. The corresponding metric is the so-called \textit{acoustic metric} (see also \cite{Unruh1981,Unruh1995} for early developments in the context of fluids) and is fully determined by the background parameters of the BEC, such as the background density or the speed of sound. In this manner, the condensate shapes the spacetime geometry experienced by the acoustic excitations. 

It has been shown that introducing suitable time-dependencies of either the scattering length or the external trapping potential allow for a one-to-one mapping of the acoustic metric onto spatially flat Friedmann-Lemaître-Robertson-Walker (FLRW) metrics \cite{Barcelo2003c,Fedichev2003,Fedichev2004,Fischer2004,Fischer2004b,Uhlmann2005,Calzetta2005,Weinfurtner2007,Weinfurtner2009,Prain2010,Bilic2013}. In this sense, all phononic modes propagate in a spacetime geometry set by the acoustic metric with the same speed of sound. In this way, modern cold atoms experiments allow for a direct analysis of quantum effects in the context of cosmological models. 

Further theoretical studies in this direction have included quantum pressure leading to a quadratic dispersion relation in the ultraviolet momentum regime and so-called rainbow FLRW metrics \cite{Weinfurtner2006,Weinfurtner2009}. Two-component BECs allowing for an additional massive phononic mode have also been investigated \cite{Liberati2006,Liberati2006b}. Recent experimental efforts for simulating an expanding universe in $d=1+1$ effective spacetime dimensions \footnote{Note that the case of $d=1+1$ spacetime dimensions is somewhat special from a cosmological point of view because the very large conformal group allows there often a map to flat space and spatial curvature is excluded.} in the laboratory can be found in \cite{Eckel2018,Wittemer2019}.

Other interesting phenomena that have been studied within the acoustic metric approach (also experimentally) comprise sonic black holes \cite{Unruh1981,Unruh1995,Visser1998,Barcelo2003,Garay2000,Garay2001}, the Unruh effect \cite{Rodriguez-Laguna2017,Hu2019,Gooding2020}, Hawking radiation \cite{Horstmann2010,Weinfurtner2011,Steinhauer2016,MunozDeNova2019} or the dynamic Casimir effect \cite{Jaskula2012}. For an overview over current experimental approaches see Ref.\ \cite{Jacquet2020}.

In this work, we focus on the analogy between the acoustic metric and the FLRW metric for an effectively $d=2+1$ dimensional isotropic trap, which is a common setup of modern BEC experiments. In contrast to previous works, we derive the acoustic metric by parameterizing the quantum fluctuations on top of the ground state in terms of the real and imaginary parts of the complex nonrelativistic scalar field, instead of magnitude and complex phase (see e.g. \cite{Floerchinger2008,Heyen2020} for a detailed discussion regarding this parameterization). 

We generalize previous studies from spatially flat to spatially curved spacetimes. To this end we explicitly consider radial dependencies of the background density profile. We find mappings between FLRW universes of positive and negative spatial curvature and the acoustic metric. 

After allowing in addition also for a time-dependent scattering length, we analyze particle production in different types of expanding cosmologies. In this sense, we extend on earlier considerations by providing an exact mapping to a more general class of cosmologies and by showing how the effect of particle production is accessible experimentally. Note that some technical details on the cosmology side are provided in our companion paper \cite{CosmologyPaper2022}.

\paragraph*{The remainder of this paper is organized as follows.} In Sec.\ \ref{sec:IsotropicTraps} we introduce isotropic traps in a quasi two-dimensional geometry and make an ansatz for the quantum effective action. We derive conditions on the background parameters such that the background density remains static. We also compute the acoustic metric from the effective action for the fluctuations and discuss the external potentials required for a one-to-one mapping between the acoustic metric and FLRW metrics of positive and negative curvature. Thereupon, we investigate the properties of the arising FLRW universes by comparing the dynamics of radially outmoving phonons. In Sec.\ \ref{sec:ParticleProduction} we derive a formalism for the spectrum of fluctuations and the two-point correlation function of a rescaled density contrast for different types of (spatially curved) cosmologies. We study both quantities in Sec.\ \ref{sec:Effects} for various experimentally accessible scenarios and point out robust features. Finally, we give a résumé and formulate an outlook in Sec.\ \ref{sec:Outlook}.

\paragraph*{Notation.} In this paper we work in SI units. For convenience, we use operator hats for creation and annihilation operators and drop them otherwise. Greek indices $\mu, \nu$ run from $0$ to $2$, while latin indices $i,j$ only run from $1$ to $2$. Also, vectors are denoted by bold symbols.

\section{Acoustic metric in 2D isotropic traps}
\label{sec:IsotropicTraps}
In the first part of this work, we discuss the acoustic metric and how it can be mapped to curved FLRW metrics for a quasi two-dimensional BEC that is confined in an isotropic trap.

\subsection{Quasi two-dimensional geometry and quantum effective action}
\label{subsec:QuantumEffectiveAction}
Let us begin our analysis with the quantum effective action of a nonrelativistic complex scalar field equipped with a quartic contact interaction term. This is an accurate description for the dynamics of a Bose-Einstein condensate, i.e. a weakly coupled Bose gas, where most atoms occupy the ground state. We denote the bosonic field expectation value (in general in the presence of sources) in $d=3+1$ spacetime dimensions by $\psi (t, \vec{r})$. We consider a pancake-type geometry in cylindrical coordinates $(r, \varphi, z)$ and a condensate that is tightly confined in the $z$-direction. Then, the extension of the condensate in $z$-direction $l_z$ is much smaller than in the longitudinal direction $l_r$, i.e. $l_z \ll l_r$, leading to a quasi $d=2+1$ dimensional geometry. Due to the strong confinement in $z$-direction, the motional degrees of freedom in this direction are frozen in, such that the mean field $\psi (t, \vec{r})$ separates according to $\psi(t, \vec{r}) = \Phi (t, r, \varphi) \zeta (z)$, where $\zeta (z)$ is typically of Gaussian form.

We study the dynamics of the field $\Phi (t, \vec{r})$ in effectively $d=2+1$ spacetime dimensions. Then, the ansatz for the action reads \cite{Floerchinger2008}
\begin{equation}
    \begin{split}
        \Gamma [\Phi] =& \int \text{d}t \, \text{d}^2r \Bigg\{i \hbar \Phi^* (\partial_t + i A_0) \Phi \\
        &- \frac{\hbar^2}{2 m} (\vec{\nabla} - i \vec{A})\Phi^* (\vec{\nabla} + i \vec{A}) \Phi - \frac{\lambda}{2} (\Phi^* \Phi)^2 \Bigg\}.
    \end{split}
    \label{eq:QuantumEffectiveAction}
\end{equation}
Here, $m$ denotes the mass of the atoms and $\lambda = \lambda(t)$ is a time-dependent coupling, which can be expressed in terms of the $s$-wave scattering length $a_s(t)$ within Born's approximation \cite{Pitaevskii2016},
\begin{equation}
    \lambda (t) = \sqrt{\frac{ 8 \pi \omega_z \hbar^{3}}{m}} a_s (t),
    \label{eq:BornApproximation}
\end{equation} 
where $\omega_z$ is the trapping frequency in $z$-direction.

Furthermore, we introduced an external $U(1)$ gauge field $A = (A_0, \vec{A})$ such that there is a symmetry of the action $\Gamma [\Phi]$ under the local $U(1)$ transformation
\begin{equation}
    \begin{split}
        \Phi (t,\vec{r}) &\to e^{- i \alpha (t, \vec{r})} \Phi(t, \vec{r}), \\
        A_0 (t, \vec{r}) &\to A_0 (t, \vec{r}) + \partial_t \alpha(t, \vec{r}), \\
        \vec{A} (t, \vec{r}) &\to \vec{A} (t, \vec{r}) + \vec{\nabla} \alpha(t, \vec{r}).
    \end{split}
\end{equation}
An external trapping potential is then given by $A_0 (t, \vec{r}) = V(t, \vec{r})/\hbar$. In our analysis, we restrict to isotropic trapping potentials of the form
\begin{equation}
    V (t, r) = \frac{m}{2} \omega^2 (t) f(r),
    \label{eq:TrappingPotential}
\end{equation}
where $\omega (t)$ is a time-dependent parameter and $f(r)$ is typically a polynomial in $r$. Without loss of generality we can assume $f(0)=0$. For example, $f(r)=r^2$ corresponds to the commonly used harmonic trap, in which case $\omega (t)$ plays the role of a trapping frequency. Moreover, in chemical equilibrium the chemical potential would enter $A_0$ such that $A_0 (t, \vec{r}) =\left( V (t, \vec{r}) - \mu\right)/\hbar$.

In the following, we will work with a linear splitting of the fundamental field $\Phi$ into a background part $\phi_0$ and a fluctuating part parametrized by two real fields $\phi_1$ and $\phi_2$, such that
\begin{equation}
    \Phi (t, \vec{r}) = \phi_0 (t, \vec{r}) + \frac{1}{\sqrt{2}} \left[ \phi_1 (t, \vec{r}) + i \phi_2 (t, \vec{r}) \right].
    \label{eq:BackgroundSplit}
\end{equation}
Therein, we allowed for general space- and time-dependencies for all fields. In the present work, we do not consider any explicit backreaction of fluctuations to the form of the action. The fluctuations are assumed to be small enough and will be kept only to linear order in equations of motion corresponding to quadratic order in the action. Conceptually this corresponds to a background field which is well described by mean field equations. We will also not consider any renormalization of the couplings in the action. In this sense, the action \eqref{eq:QuantumEffectiveAction} can actually be identified with (an approximation of) the quantum \textit{effective} action, which is renormalized already.

We are particularly interested in evaluating the effective action at the point where the field equation of motion is satisfied, 
\begin{equation}
    \frac{\delta}{\delta \Phi(t, \vec{r})} \Gamma [\Phi] = 0.
    \label{eq:QEOM}
\end{equation}
The background field $\phi_0 (t, \vec{r})$ corresponds then to an expectation value of the microscopic field or quantum operator in the absence of sources (up to a wave function renormalization constant). The normalization implicit in equation \eqref{eq:QuantumEffectiveAction} is such that $n_s=\abs{\Phi}^2$ is the superfluid density, which at vanishing temperature equals the full density $n=n_s$. For $A_0 = V/\hbar$ and $\vec{A}=0$, the classical field $\phi_0 (t, \vec{r})$ is a solution of the Gross-Pitaevskii equation  \cite{Bogoliubov1946,Lifshitz1980}
\begin{equation}
        i \hbar \partial_t \phi_0 = \left( - \frac{\hbar^2}{2m} \vec{\nabla}^2 + V + \lambda \abs{\phi_0}^2 \right) \phi_0.
    \label{eq:GrossPitaevskii2D}
\end{equation}
The superfluid behavior of the condensate mean field $\phi_0 (t, \vec{r})$ can be highlighted by introducing the Madelung representation \cite{Madelung1927},
\begin{equation}
    \phi_0 (t, \vec{r}) = \sqrt{n_0 (t, \vec{r})} e^{i S_0 (t, \vec{r})},
    \label{eq:MadelungRepresentation}
\end{equation}
with $n_0 (t, \vec{r}) = \abs{\phi_0 (t, \vec{r})}^2$ denoting the background particle number density and $S_0 (t, \vec{r})$ being the background phase of the condensate's mean field. Using the parametrization \eqref{eq:MadelungRepresentation} in Eq.\ \eqref{eq:GrossPitaevskii2D} leads to a pair of hydrodynamic equations. Namely, one obtains the local conservation law or continuity equation
\begin{equation}
    0 = \partial_t n_0 + \vec{\nabla} (n_0 \vec{v}),
    \label{eq:ContinuityEquation}
\end{equation}
and the Euler equation \footnote{Note that usually the divergence of the subsequent equation is referred to as the Euler equation.}
\begin{equation}
    0 = \hbar \partial_t S_0 + V + \lambda n_0 + \frac{\hbar^2}{2 m} (\vec{\nabla} S_0)^2.
    \label{eq:EulerEquation}
\end{equation}
We introduced the superfluid velocity via
\begin{equation}
    \vec{v} = \frac{\hbar}{m} \vec{\nabla} S_0,
    \label{eq:SuperfluidVelocity}
\end{equation}
and neglected the quantum pressure term,
\begin{equation}
\label{eq:quantumPressure}
    q = - \frac{\hbar^2}{2 m} \frac{\vec{\nabla}^2 \sqrt{n_0}}{\sqrt{n_0}},
\end{equation}
in \eqref{eq:EulerEquation}, which is the standard assumption leading to the acoustic approximation \cite{Volovik2009,Barcelo2011}. Note that $q$ in Eq.\ \eqref{eq:quantumPressure} is of second order in $\hbar$, as well as in spatial derivatives, such that it is expected to be subleading for sufficiently smooth density.

The dynamics of the background variables required to mimic FLRW universes for the fluctuating variables will be discussed in Sec.\ \ref{subsec:BackgroundDensity} and Sec.\ \ref{subsec:AcousticMetric}, respectively.

\subsection{Stationary background density profile}
\label{subsec:BackgroundDensity}
Although we have allowed for general time-dependencies of the trapping potential $V(t,r)$ and the coupling $\lambda(t)$, we are interested in describing situations where the background density $n_0$ remains static. Therefore, we do not follow the common scaling ansatz put forward in \cite{Castin1996}, but instead require the background velocity to vanish $\vec v = 0$. As a consequence, we do not need to distinguish between laboratory and comoving coordinates as they agree in static situations. The condition $\vec v = 0$ renders the continuity equation \eqref{eq:ContinuityEquation} trivial, while the Euler equation \eqref{eq:EulerEquation} evaluates to 
\begin{equation}
    0 = -  \mu_0 (t) + \frac{m}{2} \omega^2 (t) f(r) + \lambda (t) \, n_0 (r),
    \label{eq:EulerEquationStatic}
\end{equation}
where we have introduced the background chemical potential
\begin{equation}
    \mu_0 (t) = - \hbar \partial_t S_0 (t).
\end{equation}
Equation \eqref{eq:EulerEquationStatic} yields the background density profile
\begin{equation}
\begin{split}
    n_0 (r) &= \bar{n}_0 \left(1 - \frac{f(r)}{R^2} \right)\\
    &=\frac{\mu_0(t)}{\lambda(t)} - \frac{m\omega^2(t)}{2\lambda(t)}f(r). 
    \label{eq:BackgroundDensityProfile}
\end{split}
\end{equation}
Oftentimes $f(r)$ is a monotonously increasing function of radius $r$ and the condensate extends up to a radius $R$ such that $r \in [0,R]$, at which the density either drops to zero or takes a constant value. Furthermore, we introduced the constant background density at the center of the trap $\bar{n}_0 = n_0 (r=0)$, which is also the proportionality constant between the time-dependent chemical potential $\mu_0 (t)$ and the coupling $\lambda (t)$, in mean field approximation, 
\begin{equation}
    \mu_0 (t) = \bar{n}_0 \, \lambda (t).
    \label{eq:ChemicalPotentialBackgroundTimeDependence}
\end{equation}
The constant $\bar{n}_0$ is also related to the total particle number $N$ via
\begin{equation}
    N = 2 \pi \, \bar{n}_0 \int_0^R \text{d} r \, r \, \left(1 - \frac{f(r)}{R^2} \right).
    \label{eq:AtomNumber}
\end{equation}
Moreover, the size parameter $R$ appears in the proportionality constant between the time-dependent parameter $\omega(t)$ and the coupling $\lambda(t)$
\begin{equation}
    \omega^2(t) = \frac{2 \bar{n}_0}{m R^2} \, \lambda(t).
    \label{eq:TrappingFrequencyTimeDependence}
\end{equation}
If the coupling $\lambda(t)$ is changed over time, the latter condition has to be fulfilled in order to guarantee a stationary density profile of the form \eqref{eq:BackgroundDensityProfile}.

Let us discuss a couple of choices for the radial dependence of the trap encoded in $f(r)$. For $f(r)=r^2$, we obtain the well-known Thomas-Fermi density profile in a harmonic trap and $R$ corresponds to the Thomas-Fermi radius, while $f(r)= F\theta(r-R)$ with $F\to \infty$ leads to a homogeneous density profile in the region $r<R$. The latter allows for a mapping to a flat FLRW cosmology, which is discussed in detail in \cite{Weinfurtner2007}.

\begin{figure}[t!]
    \includegraphics[width=0.48\textwidth]{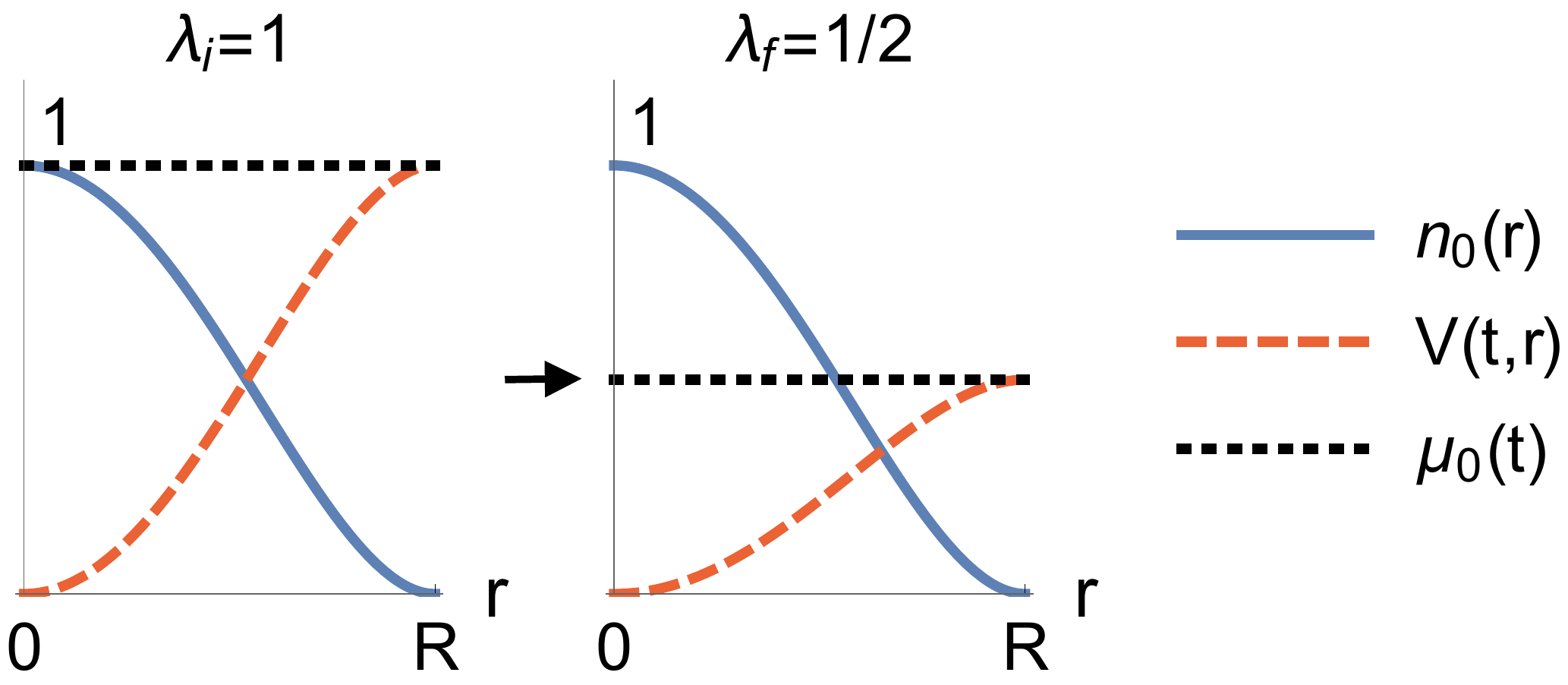}
    \caption{The density profile $n_0 (r) = \bar{n}_0 (1 - r^2/R^2)^2$ (blue solid curve) is shown together with the corresponding time-dependent trapping potential $V(t,r)$ (red dashed curve) and chemical potential $\mu_0 (t)$ (black dotted line). Note that we have set $\bar{n}_0 = 1$. Decreasing the coupling $\lambda_\text{i}=1 \to \lambda_\text{f} = \lambda_\text{i}/2 = 1/2$ and changing the trapping potential $V(t,r)$ accordingly leads to a stationary background density profile $n_0 (r)$. In order to illustrate this we set all SI units to $1$.} 
    \label{fig:DensityProfile}
\end{figure}

As we will show later, spatially curved but homogeneous and isotropic FLRW universes follow from radial dependencies of the trap and density profiles of the form
\begin{equation}
    f(r) = \pm 2 r^2 - \frac{r^4}{R^2} \hspace{0.3cm} \text{and} \hspace{0.3cm} n_0 (r) = \bar{n}_0 \left[1 \mp \frac{r^2}{R^2} \right]^2.
    \label{eq:TrapRadialDependenceExact}
\end{equation}
The time dependence of all involved quantities in the density profile of Eq.\ \eqref{eq:TrapRadialDependenceExact} (with the upper sign) is sketched in Fig.\ \ref{fig:DensityProfile}. When reducing the coupling $\lambda(t)$ over time, for example to half of its initial value, the parameter $\omega(t)$ has to be adjusted according to \eqref{eq:TrappingFrequencyTimeDependence} (the background chemical potential $\mu_0 (t)$ follows \eqref{eq:ChemicalPotentialBackgroundTimeDependence}), such that the density profile remains static. Note that decreasing the coupling $\lambda (t)$ and adjusting the parameter $\omega (t)$ accordingly over time never breaks the confinement condition $\omega(t) \ll \omega_z$, provided that it is fulfilled initially.

\subsection{Deriving the acoustic metric}
\label{subsec:AcousticMetric}
As a next step, we consider the dynamics of the fluctuations parametrized by the two real fields $\phi_1$ and $\phi_2$ introduced in Eq.\ \eqref{eq:BackgroundSplit}. To that end, we expand the effective action \eqref{eq:QuantumEffectiveAction} around the background solution $\phi_0$ to quadratic order in the fluctuating fields $\phi_1$ and $\phi_2$, which yields
\begin{equation}
    \Gamma [\Phi] = \Gamma [\phi_0] + \text{terms linear in $\phi_1, \phi_2$} + \Gamma_2 [\phi_1, \phi_2],
\end{equation}
with
\begin{widetext}
    \begin{equation}
        \begin{split}
            \Gamma_2 [\phi_1, \phi_2] =& \int \text{d}t \, \text{d}^2 r \left\{ \hbar \phi_2 \partial_t \phi_1 - \frac{\hbar^2}{4m}\left[(\vec{\nabla} \phi_1)^2 +(\vec{\nabla} \phi_2)^2  \right] -\frac{1}{2}\left(\hbar A_0 + \hbar^2 \frac{\vec{A}^2}{2m}\right)(\phi_1^2 + \phi_2^2)\right.  \\
		    & \left.- \frac{\hbar^2}{2m} \vec{A}(\phi_1 \vec{\nabla}\phi_2 -\phi_2 \vec{\nabla}\phi_1)
	        -\frac{\lambda}{2}(\phi_1,\phi_2)	\left(\begin{array}{cc} n_0 + \frac{1}{2}(\phi_0^* +\phi_0)^2 & \frac{1}{2}(\phi_0^* +\phi_0)(i\phi_0^* -i\phi_0)  \\\frac{1}{2}(\phi_0^* +\phi_0)(i\phi_0^* -i\phi_0)& n_0 + \frac{1}{2}(i\phi_0^* -i\phi_0)^2 \end{array}\right)\binom{\phi_1}{\phi_2}\right\}.
        \end{split}
        \label{eq:QEAPhiOneAndPhiTwo}
    \end{equation}
\end{widetext}
Terms linear in the fluctuating fields $\phi_1$ and $\phi_2$ cancel out at the point where the effective action is stationary (cf. Eq.\ \eqref{eq:QEOM}), therefore we only have to consider the quadratic part $\Gamma_2 [\phi_1, \phi_2]$. 

To relate the linearization for the background field \eqref{eq:BackgroundSplit} to its Madelung representation  \eqref{eq:MadelungRepresentation}, we perform a local $U(1)$ gauge transformation of the form
\begin{equation}
    \begin{split}
        \phi_0 + \frac{1}{\sqrt{2}} \left[\phi_1 + i \phi_2 \right] &\to e^{- i S_0} \left(  \phi_0 + \frac{1}{\sqrt{2}} \left[ \phi_1 + i \phi_2\right]\right), \\
        A_0 &\to A_0  + \partial_t S_0, \\ 
        \vec{A} &\to \vec{A} + \vec{\nabla} S_0.
    \end{split}
\end{equation}
This transformation redefines the fluctuating fields and rotates the -- in general complex -- background field $\phi_0$ such that it becomes real. If we again take the values before the transformation to be $A_0=V/\hbar $ and $\vec{A}=0$, then we get afterwards
\begin{equation}
    A_0 = \frac{V}{\hbar} + \partial_t S_0, \hspace{0.5cm} \text{and} \hspace{0.5cm} \vec{A}= \vec{\nabla}S_0,
\end{equation}
so that the effective action for the fluctuations becomes 
\begin{equation}
    \begin{split}
        &\Gamma_2 [\phi_1, \phi_2] \\
        &= \int \text{d} t \, \text{d}^2 r \left\{\hbar \phi_2 \partial_t \phi_1 - \frac{\hbar^2}{4m}\left[(\vec{\nabla} \phi_1)^2 +(\vec{\nabla} \phi_2)^2  \right]  \right. \\
        & -\frac{1}{2}\left(V+\hbar\partial_t S_0 + \hbar^2\frac{(\vec{\nabla}S_0)^2}{2m}\right)(\phi_1^2 + \phi_2^2)\\
		&- \frac{\hbar^2}{2m} (\vec{\nabla}S_0)(\phi_1 \vec{\nabla}\phi_2 -\phi_2 \vec{\nabla}\phi_1)\\
		& \left.  -\frac{\lambda n_0}{2}(3\phi_1^2+\phi_2^2)\right\}.
\end{split}
\end{equation}
The latter can be simplified using the Euler equation \eqref{eq:EulerEquation}, to wit
\begin{equation}
    \begin{split}
        &\Gamma_2 [\phi_1, \phi_2] \\
        &= \int \text{d} t \, \text{d}^2 r \left\{ - \frac{\hbar^2}{4m}(\vec{\nabla} \phi_2)^2  -\frac{1}{2}\phi_1\left(2\lambda n_0-\hbar^2\frac{\vec{\nabla}^2}{2m} \right)\phi_1 \right. \\
		&\left.  + \phi_1\left[-\hbar\partial_t \phi_2 -\frac{\hbar^2}{m} (\vec{\nabla}S_0) \vec{\nabla}\phi_2 -  \frac{\hbar^2}{2m}   (\vec{\nabla}^2 S_0)\phi_2\right] \right\}.
    \end{split}
\label{eq:FullFluctAction}    
\end{equation}
In the soft regime, i.e. for small momenta, one can replace $2\lambda n_0 -\hbar^2\vec{\nabla}^2/2m \to 2\lambda n_0$, which realizes the acoustic approximation for the fluctuation field. This allows one to integrate out $\phi_1$ by evaluating it on its equation of motion, leading to a quadratic effective action for $\phi_2$ only. Moreover, we also neglect the term $\vec{\nabla}^2 S_0$, that is, we assume that the background velocity is constant $\vec{v}=\text{const.}$, which is indeed fulfilled for the scenarios described in Sec.\ \ref{subsec:BackgroundDensity}, and we rescale the fluctuating field $\phi \equiv \phi_2/\sqrt{2m}$, such that it has standard mass dimension of a relativistic scalar field. Then we find
\begin{equation}
    \begin{split}
        \Gamma_2 [\phi] =& \frac{\hbar^2}{2} \int \text{d} t \, \text{d}^2 r \left\{ \frac{1}{c^2}  (\partial_t \phi)^2-  (\vec{\nabla} \phi)^2   \right. \\
		&\left. + \frac{2}{c^2}(\partial_t \phi)\, \vec{v}\cdot \vec{\nabla} \phi + \frac{1}{c^2}(\vec{v}\cdot \vec{\nabla} \phi)^2  \right\},
    \end{split}
    \label{eq:QEAPhiTwo}
\end{equation}
where we introduced the time- and space-dependent speed of sound
\begin{equation}
    c^2 (t, \vec{r}) = \frac{\lambda (t) \, n_0 (t, \vec{r})}{m}.
    \label{eq:SpeedOfSound}
\end{equation}
Finally, the latter effective action can be rewritten as an effective action for a free massless scalar field in a curved spacetime determined by the acoustic metric $g_{\mu\nu}(x)$,
\begin{equation}
    \Gamma_2 [\phi] = -\frac{\hbar^2}{2} \int \text{d} t \, \text{d}^2 r \, \sqrt{g} \, g^{\mu\nu} \partial_\mu \phi \partial_\nu \phi,
    \label{eq:QEACurvedSpacetime}
\end{equation}
where $\sqrt{g} \equiv \sqrt{- \det (g_{\mu \nu})}$. Comparing \eqref{eq:QEAPhiTwo} and \eqref{eq:QEACurvedSpacetime} reveals that the covariant components of the acoustic metric are given by
\begin{equation}
    (g^{\mu \nu}) = \begin{pmatrix}
    -1 & v^j \\
    v^i & c^2 \delta^{i j} - v^i v^j
    \end{pmatrix},
    \label{eq:CovariantAcousticMetric}    
\end{equation}
while its contravariant components read
\begin{equation}
    \left( g_{\mu \nu} \right) = \frac{1}{c^2}
    \begin{pmatrix}
    -(c^2 - v^2) & - v_j \\
    - v_i & \delta_{i j}
    \end{pmatrix}.
    \label{eq:ContravariantAcousticMetric}
\end{equation}
Furthermore, this yields $\sqrt{g}=1/c^2$.

Also, in the acoustic approximation and for a stationary background ($\vec{v}=0$) we find a simple relation between the fluctuating fields
\begin{equation}
    \phi_1 = - \frac{\hbar}{2 \lambda (t) n_0 (r)} \partial_t \phi_2,
    \label{eq:FluctuationFieldsRelation}
\end{equation}
showing that $\phi_1$ is proportional to the time derivative of $\phi_2$.

\subsection{From an acoustic metric to curved FLRW universes}
\label{subsec:MetricToFLRW}
Restricting ourselves to the scenarios described in Sec.\ \ref{subsec:BackgroundDensity}, i.e. stationary density profiles for the background density $n_0 (r)$ corresponding to $\vec{v}=0$, leads to an acoustic line element of the form
\begin{equation}
    \begin{split}
        \text{d} s^2 &= g_{\mu \nu} \text{d} x^{\mu} \text{d} x^{\nu} \\ 
        &= - \text{d}t^2 + a^2(t) \left(1 - \frac{f(r)}{R^2}\right)^{-1} (\text{d}r^2 + r^2 \text{d} \varphi^2),
    \end{split}
    \label{eq:BLineElement}
\end{equation}
where we defined a time-dependent scale factor
\begin{equation}
    a^2(t) \equiv \frac{m \, }{\bar{n}_0} \frac{1}{\lambda(t)}.
    \label{eq:ScaleFactorDefinition}
\end{equation}
In order to reshape the former into a curved FLRW line element, we continue with the particular choice $f(r) = \pm 2r^2 - r^4/R^2$ put forward in Eq.\ \eqref{eq:TrapRadialDependenceExact}, which yields the line element
\begin{equation}
    \text{d} s^2 = - \text{d}t^2 + a^2(t) \left(1 \mp \frac{r^2}{R^2}\right)^{-2} (\text{d}r^2 + r^2 \text{d} \varphi^2).
    \label{eq:LabCoordinatesLineElement}
\end{equation}
We then perform a coordinate transformation for the radial coordinate
\begin{equation}
	u(r)=\frac{r}{1 \mp \frac{r^2}{R^2}},
	\label{eq:NewRadialCoordinate}
\end{equation}
with $u \in [0, \infty)$ if we choose the negative sign, while $u \in [0, R/2]$ for the positive sign. We find the relation
\begin{equation}
	\frac{\text{d}r^2}{\left(1 \mp \frac{r^2}{R^2}\right)^2}= \frac{\text{d}u^2}{1 \pm 4\frac{u^2}{R^2}},
\end{equation}
such that the line element becomes
\begin{equation}
    \text{d} s^2 = - \text{d}t^2 + a^2(t)  \left(\frac{\text{d}u^2}{1 - \kappa u^2} + u^2 \text{d} \varphi^2\right),
    \label{eq:FLRWLineElement}
\end{equation}
which corresponds to the line element of curved FLRW universes with negative/positive spatial curvature $\kappa = \mp 4/R^2$. Therein, the size of the condensate $R$ determines the value of the scalar curvature $\kappa$, which allows the latter to be engineered in practice. Moreover, as the scale factor $a^2(t)$ is antiproportional to the coupling $\lambda (t)$ (cf. Eq.\ \eqref{eq:ScaleFactorDefinition}), decreasing (increasing) the coupling corresponds to an expanding (contracting) universe.

Interestingly, one can recover a flat FLRW universe without any additional variable transformation when realizing a homogeneous background density profile $n_0= \text{const.}$, such that the radial dependent prefactor in the spatial line element \eqref{eq:BLineElement} is absent. This is typically fulfilled in a box trap or in a sufficiently small region around the center of a (harmonic) trap.

Let us mention that for a (possibly inverted) harmonic trap $f(r)=\pm r^2$, the acoustic metric becomes
\begin{equation}
    \text{d} s^2 = - \text{d}t^2 + a^2(t) \left(1 \mp \frac{r^2}{R^2}\right)^{-1} (\text{d}r^2 + r^2 \text{d} \varphi^2),    
\end{equation}
requiring the coordinate transformation to be of the form
\begin{equation}
	u(r)=\frac{r}{\left(1 \mp \frac{r^2}{R^2}\right)^{1/2}}.
\end{equation}
Expanding the denominator of the radial differential up to quadratic order in the new radial coordinate $u$ yields
\begin{equation}
    \frac{\text{d}r^2}{1 \mp \frac{r^2}{R^2}} = \frac{\text{d}u^2}{\left(1 \pm \frac{u^2}{R^2}\right)^{2}} \approx \frac{\text{d}u^2}{1 \pm 2\frac{u^2}{R^2}},
\end{equation}
which is a reasonable approximation in a large region around the center of the trap. Then, one can also arrive at \eqref{eq:FLRWLineElement} for $\kappa = \mp 2/R^2$, such that the harmonic trap produces curved FLRW universes in a macroscopic region around the center of the trap. Typically, the approximation works well up to $r \approx 0.4R$. However, in order for the mapping to be exact within the acoustic approximation, the trapping potential has to be of the form \eqref{eq:TrapRadialDependenceExact}.

\begin{figure}[t!]
    \includegraphics[width=0.48\textwidth]{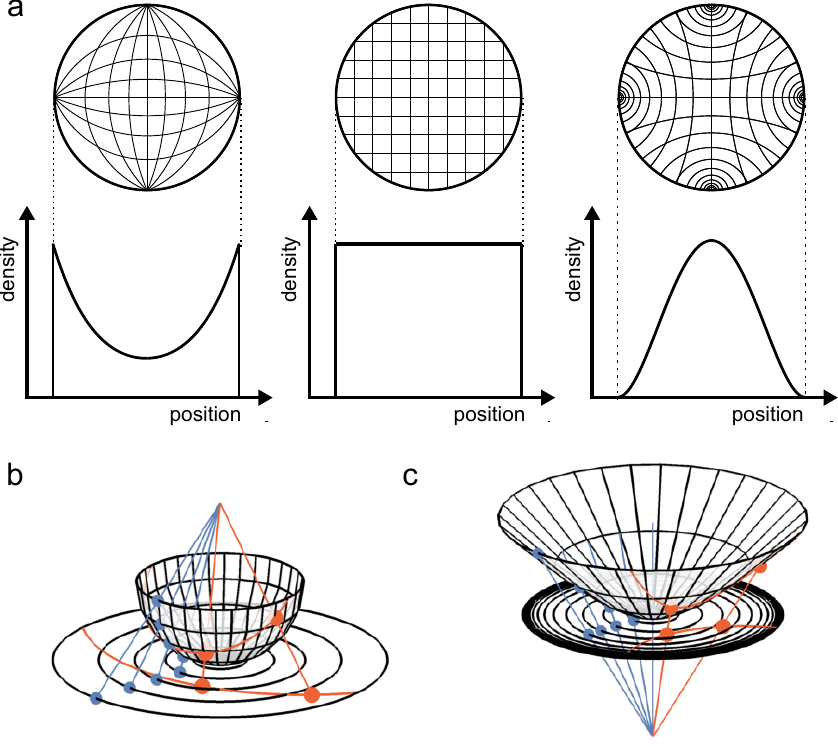}
    \caption{\textbf{a)} The one-to-one correspondence between particular radially symmetric density profiles and geometries of FLRW universes is shown, with lattices emphasizing  spatial curvature. \textbf{b)} The blue points on a positively curved universe represented as a half-sphere embedded in three-dimensional Euclidean space are projected onto a polar disk via the blue straight lines. The two red points are connected by a geodesic (red line). \textbf{c)} Similar setup for a negatively curved universe shown as a hyperboloid embedded in three-dimensional Minkowski space, which is projected onto the Poincaré disk.} 
    \label{fig:FLRWUniverses}
\end{figure}

\subsection{Models of spatially curved spacetimes}
\label{subsec:Geometries}
The relation between the density profiles \eqref{eq:TrapRadialDependenceExact} and the spatially curved FLRW universes is illustrated in Fig.\ \ref{fig:FLRWUniverses} \textbf{a)}. Thereupon, let us further comment on the different sets of spatial coordinates and their geometric meaning. 

The spatial FLRW line element \eqref{eq:FLRWLineElement} is written in terms of reduced-circumference polar coordinates in two spatial dimensions $(u,\varphi)$, which is a convenient choice also in cosmology. The radial coordinate transformation in \eqref{eq:NewRadialCoordinate} corresponds (up to a factor of $2$) to the transformation between reduced-circumference polar coordinates and polar coordinates in the polar plane $(r \le R,\varphi)$ for $\kappa > 0$, or in the Poincaré disk model $(r \le R,\varphi)$ for $\kappa < 0$. Hence, the laboratory line element \eqref{eq:LabCoordinatesLineElement} describes at every instance of time $t$, depending on spatial curvature, a polar disk or a Poincaré disk. 

The arising spatially curved geometries in the laboratory coordinates $(r,\varphi)$ may be understood from the more intuitive spherical and hyperboloid models, for which we make a distinction of cases. For the positively curved universe ($\kappa > 0$), we start from three-dimensional Euclidean space $\mathbb{R}^3$ with cartesian coordinates $(X,Y,Z)$ and line element
\begin{equation}
    \text{d} s^2 = \text{d} X^2 + \text{d} Y^2 + \text{d} Z^2.
    \label{eq:LineElementEuclidean}
\end{equation}
A two-sphere of radius $R/2$ is embedded in this space via the equation
\begin{equation}
    R^2/4 = X^2 + Y^2 + Z^2.    
    \label{eq:EmbeddingSphere}
\end{equation}
Points on the sphere can be represented by two angles ${\theta' \in [0,\pi]}$, $\varphi' \in [0,2\pi)$, which can be mapped to the global coordinates via
\begin{equation}
    (X,Y,Z) = R/2 \, \left( \sin \theta' \cos \varphi', \sin \theta' \sin \varphi', \cos \varphi' \right).
\end{equation}
The induced metric on the two-sphere reads
\begin{equation}
    \text{d}s^2 = R^2 / 4 \left( \text{d} \theta'^2 + \sin^2 \theta' \, \text{d} \varphi'^2 \right),
\end{equation}
which corresponds to an intuitive representation of a positively curved space (cf. Fig.\ \ref{fig:FLRWUniverses} \textbf{b})). However, the two-sphere can also be mapped to the laboratory picture, i.e. the polar disk with coordinates $(r, \varphi)$, via a stereographic projection from the north pole $(0,0,R/2)$ onto the disk located at the south pole $(0,0,-R/2)$ (illustrated with blue lines connecting points on the sphere and in the plane in Fig.\ \ref{fig:FLRWUniverses} \textbf{b})). Then, the laboratory coordinates are related to the coordinates in $S^2$ by
\begin{equation}
    (r, \varphi) = \left(R \cot \frac{\theta'}{2}, \varphi' \right),
\end{equation}
and we obtain the spatial part of the FLRW line element in the laboratory \eqref{eq:LabCoordinatesLineElement}, with positive sign.

In case of the negatively curved universe ($\kappa < 0$), we have to start from three-dimensional Minkowski space $\mathbb{M}^3$ instead \cite{Balazs1986}. Adapting the cartesian coordinates $(X,Y,Z)$ from before leads to a line element of the form 
\begin{equation}
    \text{d} s^2 = \text{d} X^2 + \text{d} Y^2 - \text{d} Z^2,
    \label{eq:LineElementMinkwoski}
\end{equation}
with an additional minus sign compared to \eqref{eq:LineElementEuclidean}. Then, the upper hyperboloid can be embedded in this space through (cf. Fig.\ \ref{fig:FLRWUniverses} \textbf{c}))
\begin{equation}
    -R^2/4 = X^2 + Y^2 - Z^2
    \label{eq:EmbeddingHyperboloid}
\end{equation}
with $Z > 0$. On the hyperboloid, we choose a pseudoangle $\sigma' \in [0,\infty)$ instead of an angle $\theta' \in [0,\pi]$, but keep the azimuthal angle ${\varphi' \in [0,2\pi)}$. We can express the global coordinates in terms of the latter coordinates as
\begin{equation}
    (X,Y,Z) = R/2 \, \left(\sinh \sigma' \cos \varphi', \sinh \sigma' \sin \varphi', \cosh \sigma' \right),
\end{equation}
leading to an induced metric
\begin{equation}
    \text{d}s^2 = R^2/4 \, \left(\text{d} \sigma'^2 + \sinh^2 \sigma' \, \text{d} \varphi' \right),
\end{equation}
on the hyperboloid. Finally, we recover the laboratory coordinates $(r, \varphi)$ by a projection of the hyperboloid onto the Poincaré disk located at the south pole of the hyberboloid, i.e. at $(0,0,R/2)$, using the apex $(0,0,-R/2)$ of the lower hyperboloid (not shown in Fig.\ \ref{fig:FLRWUniverses} \textbf{c)}) as the base point. This projection is sketched with blue straight lines in Fig.\ \ref{fig:FLRWUniverses} \textbf{c)}. We obtain the relation
\begin{equation}
    (r, \varphi) = \left(R \coth \frac{\sigma'}{2}, \varphi' \right),
\end{equation}
and the spatial part of the FLRW line element \eqref{eq:LabCoordinatesLineElement} with a minus sign for the metric in the Poincaré disk.

\begin{figure*}[t!]
    \centering
    \includegraphics[width=0.27\textwidth]{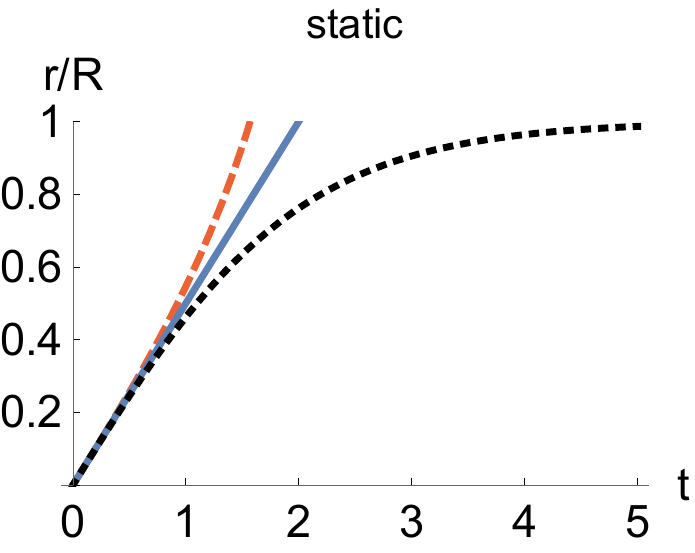}
    \includegraphics[width=0.27\textwidth]{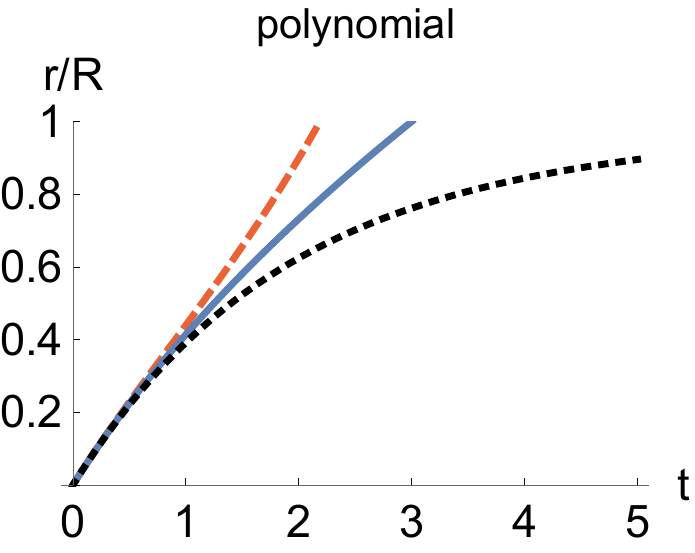}
    \includegraphics[width=0.38\textwidth]{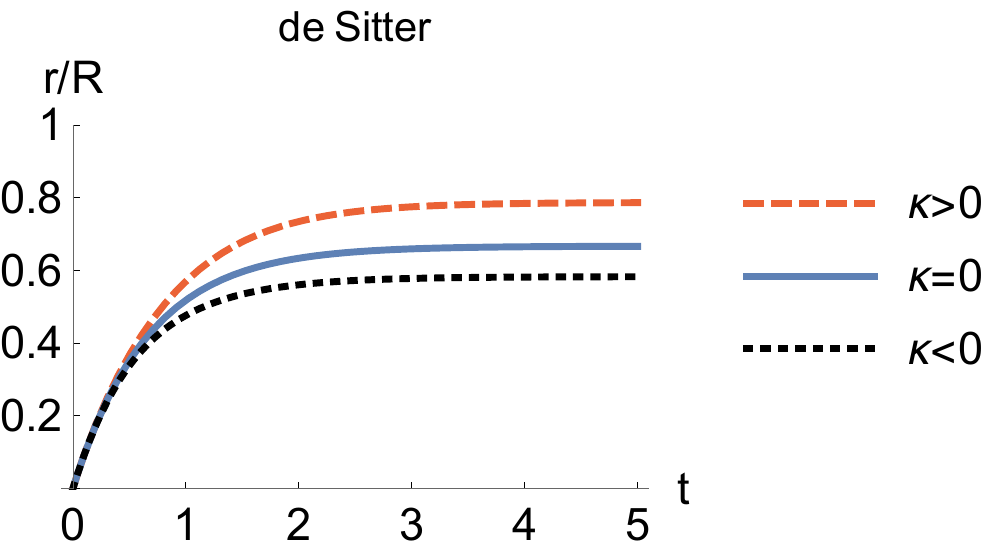}
    \caption{The trajectories of radially outmoving phonons for the trap profiles given in Eq.\ \eqref{eq:TrapShape} are shown for a static situation (left panel, with values $Q=2, t_0=-1, R=1$), a polynomial scale factor $a(t)=Q(t-t_0)^ {1/2}$ (middle panel, with values $Q=2, t_0=-1, R=1$) and a de Sitter type scale factor (right panel, with values $a_0=1, H=1.5, R=1$). Clearly, the spatial curvature influences the bending of the curves, with stronger / weaker bending for open (black dotted line) / closed (red dashed line) universes compared to the flat one (blue solid line). Again, for convenience, we work with dimensionless parameters.}
    \label{fig:PhononTrajectories}
\end{figure*}

\subsection{Phonon trajectories}
\label{subsec:PhononTrajectories}
To exemplify the influence of spatial curvature on the dynamics of acoustic excitations in the Bose-Einstein condensate, we consider the motion of a radially outmoving wave packet starting from the center of the trap. Phonons follow null geodesics in an acoustic spacetime, so we look for trajectories with $\text{d} s^2 = 0.$

 We proceed with the general line element \eqref{eq:BLineElement}, which is expressed in terms of the radial coordinate in the laboratory $r$. For radial geodesics we have $ \text{d} \varphi = 0$; this, together with $\text{d} s^2 = 0$, leads to the simple differential equation
\begin{equation}
    \frac{\text{d}t}{a(t)} = \frac{\text{d}r}{\sqrt{1 - f(r)/R^2}},
    \label{eq:TrajectoriesEoM}
\end{equation}
with the initial condition $r(t=0)=0$. The three different types of spatial curvature are generated by
\begin{equation}
    f(r) = \begin{cases} 
    -2 r^2 - r^4/R^2  & \text{for} \hspace{0.3cm} \kappa > 0, \\
    0 & \text{for} \hspace{0.3cm} \kappa = 0, \\
    +2 r^2 - r^4/R^2  & \text{for} \hspace{0.3cm} \kappa < 0,
    \end{cases}
    \label{eq:TrapShape}
\end{equation}
so that to the equation of motion \eqref{eq:TrajectoriesEoM} has general solutions of the form
\begin{equation}
    r(t) = R \, \begin{cases}
    \tan z(t)  & \text{for} \hspace{0.3cm} \kappa > 0, \\
    z(t) & \text{for} \hspace{0.3cm} \kappa = 0, \\
    \tanh z(t) & \text{for} \hspace{0.3cm} \kappa < 0,
    \end{cases}
    \label{eq:TrajectoriesSolution}
\end{equation}
where the argument $z(t)$ can be obtained from the scale factor $a(t)$ via
\begin{equation}
    z(t) = \frac{1}{R} \, \int_0^t \frac{\text{d} t^{\prime}}{a(t^{\prime})}.
\end{equation}
For a quadratic trap profile, $f(r)=\pm r^2$, we would obtain $\sin \, (\sinh)$ instead of $\tanh \, (\tan)$ in Eq.\ \eqref{eq:TrajectoriesSolution}.

It is of particular interest to consider polynomial scale factors. To explicitly allow for expanding as well as contracting scenarios, we write
\begin{equation}
    a(t) = Q \, \abs{t - t_0}^{\gamma},
    \label{eq:ScaleFactorPolynomial}
\end{equation}
where $Q= a(t=0) \, \abs{t_0}^{- \gamma} > 0$ and $t_0$ are free parameters to be tuned in experiments. The latter family of scale factors comprises the analogs of radiation dominated ($\gamma=2/3$) and matter dominated ($\gamma=1$) universes. Note that the units of the parameter $Q$ depend on the power $\gamma$.

For the polynomial scale factors with $\gamma \neq 1$ we find
\begin{equation}
    z(t) = \frac{1}{Q R} \frac{\text{sgn}(t - t_0)\abs{t - t_0}^{1-\gamma} + \text{sgn}(t_0)\abs{ t_0}^{1-\gamma}}{1-\gamma},
\end{equation}
which is shown in the middle panel of Fig.\ \ref{fig:PhononTrajectories} for $\gamma=1/2$. For the specific case $\gamma = 0$ corresponding to a static situation (cf. left panel in Fig.\ \ref{fig:PhononTrajectories}), we get
\begin{equation}
    z(t) = \frac{t}{Q R},
\end{equation}
while $\gamma = 1$ corresponds to
\begin{equation}
    z(t) = \frac{1}{Q R} \ln \frac{\abs{t-t_0}}{\abs{t_0}}.
\end{equation}

Another interesting example is the de Sitter universe, which is characterized by the scale factor
\begin{equation}
    a(t) = a_0 \, e^{H t},
    \label{eq:ScaleFactordeSitter}
\end{equation}
where $H$ denotes the Hubble parameter and $a_0 = a (t=0)$ is now the initial scale factor. In this case we find
\begin{equation}
    z(t) = \frac{1}{a_0 H R} \left(1 - e^{- H t} \right ).
\end{equation}
The resulting trajectories are depicted in the right panel of Fig.\ \ref{fig:PhononTrajectories}. The quantitative influence of the spatial curvature parameter $\kappa$ is clearly visible.

\section{Particle Production}
\label{sec:ParticleProduction}
We now turn to the phenomenon of particle production, which arises when the spacetime geometry becomes time-dependent. We develop a formalism to access particle production experimentally within the FLRW cosmology paradigm and consider all three types of spatial curvature engineered via \eqref{eq:TrapShape}.

\subsection{Klein-Gordon equation and mode functions}
\label{subsec:ModeFunctions}
Let us start with the action \eqref{eq:QEACurvedSpacetime} for the fluctuation field $\phi$. Varying the latter using Eq.\ \eqref{eq:FLRWLineElement} leads to a Klein-Gordon equation
\begin{equation}
    \begin{split}
        0 &= \partial_\mu \left(\sqrt{g} \, g^{\mu \nu} \, \partial_\nu \phi \right) \\
        &= 2 a(t)\dot{a}(t)\dot{\phi} + a^2(t) \ddot{\phi} - \Delta \phi,     
    \end{split}
    \label{eq:KleinGordonFLRW}
\end{equation}
where $\dot{f} \equiv \partial_t f$ denotes the partial derivative with respect to time $t$. The exact form of the Laplace-Beltrami operator depends on the spatial curvature (see below).

For spatially curved universes with $\kappa > 0$, we transform the radial coordinate $u$ to an angle $\theta$. It is convenient to extend from the half sphere to the full sphere such that $\theta \in [0,\pi)$. Similarly, for $\kappa<0$ we can introduce the pseudoangle $\sigma\in[0,\infty)$ and for $\kappa=0$ we work with an infinitely extended disk, $u\in[0,\infty)$. In summary we work with a radial coordinate $\theta, u$ or $\sigma$ defined by
\begin{equation}
    u = \begin{cases}
    \frac{\sin \theta}{\sqrt{\abs{\kappa}}} &\text{for}\hspace{0.3cm} \kappa > 0, \\
    u &\text{for}\hspace{0.3cm} \kappa = 0, \\
    \frac{\sinh \sigma}{\sqrt{\abs{\kappa}}} &\text{for}\hspace{0.3cm} \kappa < 0.
    \end{cases}
    \label{eq:u}
\end{equation}

In these coordinates, one has
\begin{equation}
\begin{split}
    \sqrt{g} = a^2(t)\times \begin{cases}
    \frac{\sin\theta}{\abs{\kappa}} &\text{for}\hspace{0.2cm} \kappa > 0, \\
    u &\text{for}\hspace{0.2cm} \kappa = 0, \\
    \frac{\sinh\sigma}{\abs{\kappa}} &\text{for}\hspace{0.2cm} \kappa < 0,
    \end{cases}
\end{split}
\end{equation}
and the isotropic Laplace-Beltrami operator in Eq.\ \eqref{eq:KleinGordonFLRW} takes the simple form \cite{Ratra1995,Ratra2017,Argyres1989}
\begin{equation}
    \Delta = \begin{cases}
    \abs{\kappa} \Big[ \frac{
    1}{\sin \theta}\partial_{\theta}\left(\sin \theta \, \partial_{\theta} \right) + \frac{
    1}{\sin^2\theta}\partial_{\varphi}^2 \Big] &\text{for}\hspace{0.2cm} \kappa > 0, \\
    \partial^2_u + \frac{1}{u} \partial_u + \frac{1}{u^2} \partial_{\varphi}^2 &\text{for}\hspace{0.2cm} \kappa = 0, \\
    \abs{\kappa} \Big[\frac{
    1}{\sinh \sigma}\partial_{\sigma}\left(\sinh \sigma \, \partial_{\sigma} \right) + \frac{
    1}{\sinh^2\sigma}\partial_{\varphi}^2 \Big] &\text{for}\hspace{0.2cm} \kappa < 0.
    \end{cases}
    \label{eq:LaplaceOperator}
\end{equation}
The Laplace-Beltrami operator can be diagonalized through an eigenvalue equation
\begin{equation}
    \Delta \mathcal{H}_{km} (u,\varphi) = h (k) \, \mathcal{H}_{km} (u,\varphi),
    \label{eq:LaplaceEigenvalueEquation}
\end{equation}
where we have introduced the radial wave numbers $k$ and $l$, which are related by $k=\sqrt{\abs{\kappa}}l$, and the azimuthal wave number $m$, with the ranges
\begin{equation}
\begin{split}
     &l \in \mathbb{N}_0, m \in \{-l,...,l \} \hspace{0.3cm} \text{for} \hspace{0.2cm} \kappa > 0, \\
    &k \in \mathbb{R}^+_0, m \in \mathbb{Z} \hspace{1.36cm} \text{for} \hspace{0.2cm} \kappa = 0, \\
    &l \in \mathbb{R}^+_0, m \in \mathbb{Z} \hspace{1.44cm} \text{for} \hspace{0.2cm} \kappa < 0,
    \label{eq:MomentumVector}
\end{split}
\end{equation}
along with sets of complete and orthonormal eigenfunctions
\begin{equation}
    \mathcal{H}_{km} (u,\varphi) = \begin{cases}
    Y_{lm}(\theta,\varphi) &\text{for} \hspace{0.2cm} \kappa > 0, \\
    X_{km} (u, \varphi) &\text{for} \hspace{0.2cm} \kappa = 0, \\
    W_{lm} (\sigma, \varphi) &\text{for} \hspace{0.2cm} \kappa < 0.
    \end{cases}
    \label{eq:LaplaceEigenfunctions}
\end{equation}
Here, $Y_{lm}(\theta,\varphi)$ are an adapted version of the spherical harmonics,
\begin{equation}
    Y_{lm}(\theta,\varphi) = \sqrt{\frac{(l-m)!}{(l+m)!}} \, e^{im\varphi} \, P_{lm}(\cos \theta),
    \label{eq:SphericalHarmonics}
\end{equation}
with $P_{lm}(\cos\theta) = (-1)^mP^m_l(\cos\theta)$ denoting the associated Legendre polynomials. Besides this sign, the choice in \eqref{eq:SphericalHarmonics} differs by a factor $\sqrt{4\pi}/\sqrt{2l+1}$ from the standard definition of the spherical harmonics. For the flat two-dimensional space one may use polar waves defined in terms of Bessel functions of the first kind
\begin{equation}
    X_{km} (u, \varphi) = e^{i m \varphi} \, J_m (k u) .
\label{eq:EigenfunctionFlat}
\end{equation}
Finally, $W_{lm} (\sigma, \varphi)$ are the eigenfunctions for $\kappa < 0$, which are given by
\begin{equation}
    \begin{split}
        W_{lm} (\sigma, \varphi) = &(-i)^m \frac{\Gamma(il+1/2)}{\Gamma(il+m+1/2)} \\
        &\times e^{im\varphi} P^m_{il-1/2}\left(\cosh \sigma\right),
    \label{eq:EigenfunctionOpen}
    \end{split}
\end{equation}
wherein $P^m_{il-1/2}\left(\cosh \sigma\right)$ are conical functions corresponding to analytically continued Legendre functions. The functions in equations \eqref{eq:SphericalHarmonics}, \eqref{eq:EigenfunctionFlat} and \eqref{eq:EigenfunctionOpen} are normalized with respect to a scalar product as discussed in detail in \cite{CosmologyPaper2022}.

With these conventions, the eigenvalues $h(k)$ of the Laplace-Beltrami operator $\Delta$ defined in Eq.\ \eqref{eq:LaplaceOperator} read
\begin{equation}
    h(k) = \begin{cases}
    - k (k+\sqrt{\abs{\kappa}}) & \text{for}\hspace{0.3cm} \kappa > 0, \\
    - k^2 & \text{for}\hspace{0.3cm} \kappa = 0, \\
    - \left( k^2 + \frac{1}{4}\abs{\kappa}\right) & \text{for}\hspace{0.3cm} \kappa < 0,
    \end{cases}
    \label{eq:LaplaceEigenvalues}
\end{equation}
with the relation $k = \sqrt{\abs{\kappa}} l$ understood in the two spatially curved cases. Note that $h(k=0)$ is only nonvanishing for the open universe, $\kappa<0$, and $\sqrt{|\kappa|}/2$ acts there like an effective mass gap.

The fluctuation field $\phi$ is quantized as usual such that it obeys the (equal time) bosonic commutation relations,
\begin{equation}
    \begin{split}
        &[\phi(t, u,\varphi), \pi(t,u^{\prime},\varphi^{\prime})] \\
        &= i \hbar \delta(\varphi-\varphi^{\prime})\times
        \begin{cases}
        \delta(\theta-\theta') & \text{for}\hspace{0.3cm} \kappa > 0,\\
        \delta(u-u')& \text{for}\hspace{0.3cm} \kappa = 0,\\
        \delta(\sigma-\sigma')& \text{for}\hspace{0.3cm} \kappa < 0,
        \end{cases}
    \end{split}
    \label{eq:FLRWCommutationRelationsFields}
\end{equation}
where
\begin{equation}
    \begin{split}
        \pi(t,u,\varphi) = \frac{\delta \Gamma_2 [\phi]}{\delta \dot{\phi}} &= \hbar^2 \sqrt{g} \dot{\phi},
    \end{split}
\end{equation}
denotes the conjugate momentum field. 

To solve the linearized equation of motion \eqref{eq:KleinGordonFLRW} for the different classes of universes, we expand the quantum field $\phi$ in terms of the corresponding eigenfunctions of the Laplace-Beltrami operator \eqref{eq:LaplaceEigenfunctions},
\begin{equation}
    \begin{split}
        &\phi(t, u,\varphi)\\ 
        &= \int_{k,m}\left[\hat{a}_{km} \mathcal{H}_{km} (u,\varphi) v_k (t) + \hat{a}^{\dagger}_{km} \mathcal{H}_{km}^{*} (u,\varphi) v_k^* (t) \right],
    \end{split}
    \label{eq:ModeExpansion}
\end{equation}
and similar for the conjugate momentum field $\pi(t, u,\varphi)$. Therein, we used the abbreviation
\begin{equation}
    \int_{k,m}  = \begin{cases}
    \sum_{l=0}^{\infty} \abs{\kappa}\frac{l+1/2}{2\pi} \sum_{m=-l}^{l} &\text{for} \hspace{0.2cm} \kappa > 0, \\
    \int \frac{\text{d}k}{2\pi} \, k \sum_{m=-\infty}^{\infty} &\text{for} \hspace{0.2cm} \kappa = 0,\\
    \int \frac{\text{d}l}{2\pi} \,\abs{\kappa} l \tanh(\pi l) \sum_{m=-\infty}^{\infty} &\text{for} \hspace{0.2cm} \kappa < 0,
    \end{cases}
\label{eq:MomentumIntegral}
\end{equation}
for the two-dimensional momentum integral. Also, we have introduced creation and annihilation operators $\hat{a}^{\dagger}_{km}$ and $\hat{a}_{km}$, respectively, which fulfill the bosonic commutation relations
\begin{equation}
    \begin{split}
        &[\hat{a}_{km}^{\dagger},\hat{a}_{k^{\prime}m^{\prime}}^{\dagger}] = [\hat{a}_{km},\hat{a}_{k^{\prime}m^{\prime}}] = 0, \\
        &[\hat{a}_{km},\hat{a}_{k^{\prime}m^{\prime}}^{\dagger}] = 2\pi \delta_{mm^{\prime}} \begin{cases}
        \frac{\delta_{ll'}}{\abs{\kappa}(l+1/2)} & \text{for}\hspace{0.3cm} \kappa > 0, \\
          \frac{\delta(k-k^{\prime})}{k} & \text{for}\hspace{0.3cm} \kappa = 0,  \\
          \frac{\delta(l-l')}{\abs{\kappa}l \tanh(\pi l)} & \text{for}\hspace{0.3cm} \kappa < 0, 
        \end{cases}
    \end{split}
    \label{eq:CommutationRelationsBosonicOperators}
\end{equation}
\iffalse
We use here the abbreviation
\begin{equation}
    \int_{k,m}  = \begin{cases}
    \sum_{l=0}^{\infty} \abs{\kappa}\frac{l+1/2}{2\pi} \sum_{m=-l}^{l} &\text{for} \hspace{0.2cm} \kappa > 0, \\
    \int \frac{\text{d}k}{2\pi} \, k \sum_{m=-\infty}^{\infty} &\text{for} \hspace{0.2cm} \kappa \leq 0.
    \end{cases}
\end{equation}
We have introduced creation and annihilation operators $\hat{a}^{\dagger}_{km}$ and $\hat{a}_{km}$, respectively, which fulfill the bosonic commutation relations
\begin{equation}
    \begin{split}
        &[\hat{a}_{km}^{\dagger},\hat{a}_{k^{\prime}m^{\prime}}^{\dagger}] = [\hat{a}_{km},\hat{a}_{k^{\prime}m^{\prime}}] = 0, \\
        &[\hat{a}_{km},\hat{a}_{k^{\prime}m^{\prime}}^{\dagger}] = 2\pi \delta_{mm^{\prime}} \times \begin{cases}
            \frac{\delta_{lL}}{\abs{\kappa}(l+1/2)} & \text{for}\hspace{0.3cm} \kappa > 0, \\
            (\abs{\kappa} + \delta_{\kappa0})\frac{\delta(k-k^{\prime})}{k} & \text{for}\hspace{0.3cm} \kappa \leq 0,  
        \end{cases}
        \label{eq:CommutationRelationsBosonicOperators}
    \end{split}
\end{equation}
\fi
together with the time-dependent mode functions $v_{k} (t)$ and $v_k^{*} (t)$.

Combining the latter statements, we find from the Klein-Gordon equation \eqref{eq:KleinGordonFLRW} the so-called mode equation
\begin{equation}
    \ddot{v}_k (t) + 2\frac{\dot{a}(t)}{a(t)}\dot{v}_k(t) - \frac{h(k)}{a^2(t)} v_k(t) = 0.
    \label{eq:ModeEquation}
\end{equation}
For given initial conditions at some point in time, one can determine $v_k(t)$ by solving Eq.\ \eqref{eq:ModeEquation}. Note that the influence of the spatial curvature $\kappa$ is fully encoded in $h(k)$. On the scales relevant for typical experiments, i.e. for $R \sim 10^{-5}$ m, $h(k)$ turns out to be practically independent of the spatial curvature $\kappa$ for $k/\sqrt{\bar{n}_0} \gtrsim  0.1$.

The mode functions are further constrained as a result of the canonical commutation relation \eqref{eq:FLRWCommutationRelationsFields}, the orthonormality properties of the functions $\mathcal{H}_{km} (u,\varphi)$ (see \cite{CosmologyPaper2022}), and the bosonic commutation relations fulfilled by the creation and annihilation operators \eqref{eq:CommutationRelationsBosonicOperators}, leading to a normalization condition in terms of the Wronskian,
\begin{equation} 
     \text{Wr}[v_k, v_k^*] = a^2(t) \hbar \left[  v_k \dot{v}_{k}^{*} - \dot{v}_{k} v_k^{*}  \right] = i.
     \label{eq:ModeRelation}
\end{equation}
By using \eqref{eq:ModeEquation} one can show that \eqref{eq:ModeRelation} is fulfilled at all times when it is fulfilled at one point in time.

Let us note here that \eqref{eq:ModeEquation} is a second order differential equation, and \eqref{eq:ModeRelation} is just a single constraint, so that the mode functions $v_k(t)$ are not fixed completely. In fact, different choices of mode functions correspond to different choices of creation and annihilation operators, and they are related by Bogoliubov transformations. Note that the quantum field $\phi$ in \eqref{eq:ModeExpansion} is independent of this choice. This is important to understand particle production, and will be discussed next.

\subsection{Bogoliubov transformations}
\label{subsec:BogoliubovTrafo}

Let us start by considering a situation (in the following called region I) where the scale factor is constant, $a(t)=a_\text{i}$. In that case one has a preferred choice of mode functions $v_k(t)$ given by oscillatory modes, 
\begin{equation}
    v_k^{\text{I}} (t) = \frac{\exp (-i \omega_k^{\text{I}} t)}{a_\text{i} \sqrt{2\hbar \omega_k^{\text{I}}}},
    \label{eq:planewavetime}
\end{equation}
with positive frequency $\omega_k^{\text{I}} = \sqrt{-h(k)}/a_\text{i}$. Associated with this choice of mode functions are operators $\hat a_{km}$ and $\hat a_{km}^\dagger$ that annihilate and create the corresponding phonons, and a ``vacuum'' state $\ket{\Omega}$ that has no such excitations,
\begin{equation}
    \hat a_{km}\ket{\Omega} = 0.
    \label{eq:defOmegaState}
\end{equation}
The state $\ket{\Omega}$ describes the ground state of a (weakly interacting) Bose-Einstein condensate with no excitations. More generally, one may also take some other state as a starting point, for example with fixed temperature $T$. 

Let us now assume that at some time $t_\text{i}$ the scale factor $a(t)$ becomes time dependent, until it becomes constant again at time $t_\text{f}$. For the intermediate times $t_\text{i} < t < t_\text{f}$ (called region II in the following), the mode functions are determined as solutions of equation \eqref{eq:ModeEquation}, with initial conditions at $t=t_\text{i}$ set by continuity to the solution \eqref{eq:planewavetime}. We stress that for a time-dependent scale factor (in region II), the solution $v_k(t)$ obtained in this way will not be of the oscillatory form \eqref{eq:planewavetime}.  As a consequence, the notions of vacuum states and (quasi)particles become more involved. Mathematically, this is related to the absence of a global timelike Killing vector field in that region, which would allow to define positive frequency waves.

\begin{figure}[t!]
    \includegraphics[width=0.48\textwidth]{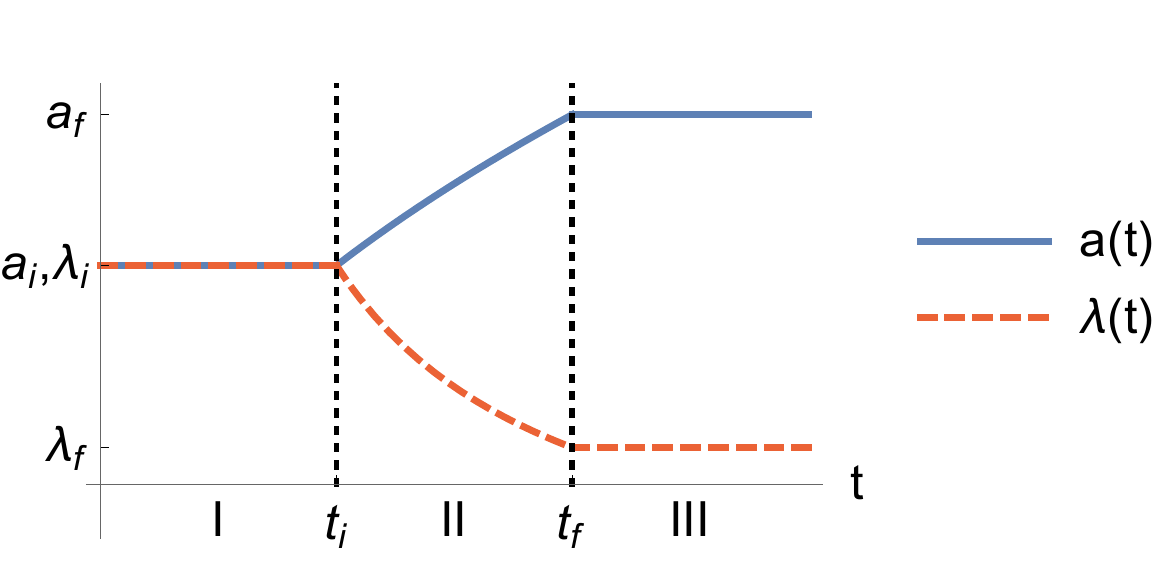}
    \caption{The time dependence of the scale factor $a(t)$ (blue solid line) and the coupling $\lambda(t)$ (red dashed line) are shown for the three regions I, II and III. Up to $t_\text{i}$, both are held constant, while they become time dependent in region II (the plot above correspond to a polynomial expansion with $\gamma = 1/2$, $Q = m/\bar{n}_0 = 1$ and $t_0=0$, setting all units to $1$) this time dependence leads to particle production, observed in the stationary region III for $t > t_\text{f}$.} 
    \label{fig:Expansion}
\end{figure}

We concentrate on an experimental procedure sketched in Fig.\ \ref{fig:Expansion}. Keeping in mind that the scale factor $a(t)$ is controlled by the inverse coupling $\lambda (t)$ according to \eqref{eq:ScaleFactorDefinition}, one may first engineer certain initial values $a_\text{i}$ and $\lambda_\text{i}$ in a time interval I up to $t_\text{i}$. Then, the coupling $\lambda(t)$ is varied over time in time interval II, which simulates a FLRW universe with time-dependent scale factor $a(t)$. Finally, the variation is stopped at $t_\text{f} = t_{\text{i}} + \Delta t$, corresponding to a stationary FLRW universe with scale factor $a_\text{f}$ in time interval III.

Let us then consider times $t>t_\text{f}$ where we assume that the scale factor is again constant, $a(t)=a_\text{f}$ (region III). Here one can again find solutions of Eq.\ \eqref{eq:ModeEquation} in terms of modes with positive frequencies,
\begin{equation}
    u_k^{\text{III}} (t) = \frac{\exp (-i \omega_k^{\text{III}} t)}{a_\text{f} \sqrt{2\hbar \omega_k^{\text{III}}}},
    \label{eq:uPlaneWaves}
\end{equation} 
where now $\omega_k^{\text{III}} = \sqrt{-h(k)}/a_\text{f}$. Associated to these modes are operators $\hat b_{km}$ and $\hat b^\dagger_{km}$ and a corresponding ``vacuum'' state $\ket{\Psi}$, such that
\begin{equation}
\hat b_{km} \ket{\Psi} = 0.
\end{equation}

It is now important to note that the mode functions $v_k(t)$ obtained from extending the solution in region I to region II and then into region III will actually be a linear superposition of positive and negative frequency solutions, so that one can write
\begin{equation}
    u_k =\alpha_k v_k + \beta_{k} v_k^{*}, \quad\quad v_k = \alpha^*_k u_k - \beta_k u^*_k, 
    \label{eq:BogoliubovTransformationModes}
\end{equation}
with the complex-valued and time-independent Bogoliubov coefficients $\alpha_k$ and $\beta_k$. From the normalization condition \eqref{eq:ModeRelation} applied to $u_k(t)$ follows that the coefficients need to satisfy $\abs{\alpha_k}^2 - \abs{\beta_k}^2 = 1$. In terms of the Wronskian defined in \eqref{eq:ModeRelation} one has
\begin{equation}
\alpha_k = \text{Wr}[u_k, v_k^*]/i, \quad\quad \beta_k = -\text{Wr}[u_k, v_k]/i.
\label{eq:BogoliubovInTermsOfWroskian}
\end{equation}

For completeness, we also mention the relation between the two sets of creation and annihilation operators,
\begin{equation}
    \hat{b}_{km} =  \alpha_k^* \hat{a}_{km} - \beta_k^* (-1)^m \hat{a}^{\dagger}_{k,-m}.
    \label{eq:BogoliubovTransformationOperators}
\end{equation}
In particular it follows that the state $\ket{\Omega}$, which is initially a vacuum state in the sense of Eq.\ \eqref{eq:defOmegaState}, is not an empty state with respect to the excitations annihilated by the operators $\hat b_{km}$. This is the essence of particle production due to a time dependent scale factor $a(t)$.

The task is then to solve the mode equation \eqref{eq:ModeEquation} in all three regions I, II and III and to identify the Bogoliubov coefficients $\alpha_k$ and $\beta_k$ through Eq.\ \eqref{eq:BogoliubovInTermsOfWroskian}, from which all following quantities can be derived.

\subsection{Rescaled density contrast: correlation function and spectrum}
\label{subsec:CorrelationFunctionAndSpectrum}
To access the phenomenon of particle production experimentally, we introduce a rescaled density contrast
\begin{equation}
    \delta_c (t, u, \varphi) = \sqrt{\frac{n_0 (u)}{\bar{n}_0^3}} \left[n (t, u, \varphi) - n_0 (u)\right],
    \label{eq:RescaledDensityContrast}
\end{equation}
where $n (t,u,\varphi) = \abs{\Phi(t,u,\varphi)}^2$ denotes the full condensate density, and $\bar n_0$ is the density in the center of the trap. In this way, the rescaled density contrast is dimensionless, and using Eqs.\ \eqref{eq:BackgroundSplit} as well as \eqref{eq:FluctuationFieldsRelation} one has to leading order $\delta_c \sim \partial_t \phi$.
%with a particular choice for the prefactor justified below. 
Note here that for a box potential the given prefactor is constant, while for other trapping potentials, such as those put forward in Eq.\ \eqref{eq:TrapRadialDependenceExact}, it has a substantial dependence on $u$ in the outer regions of the trap. 

In the following, we consider the equal time two-point correlation function of the rescaled density contrast after the expansion has ceased, $t \ge t_{\text{f}}$, i.e.
\begin{align}
    \mathcal{G}_{n n} (t; u, u', \varphi, \varphi') = \braket{\delta_c (t,u,\varphi) \delta_c (t,u',\varphi')},
    \label{eq:DensityDensityCorrelatorDefinition}
\end{align}
which is a typical observable in modern ultracold atom experiments. One can show that in leading order in fluctuating fields, where we can assume $\braket{n(t, u, \varphi)} = n_0 (u)$, the latter is proportional to the connected two-point correlation function of time derivatives of fields
\begin{equation}
    \mathcal{G}_{n n} (t; u, u', \varphi, \varphi') = \frac{\hbar^2 m}{\lambda^2_{\text{f}} \bar{n}_0^3} \, \mathcal{G}_{\dot{\phi} \dot{\phi}} (t, L),
    \label{eq:DensityDensityCorrelatorLeadingOrder}
\end{equation}
where
\begin{equation}
   \mathcal{G}_{\dot{\phi} \dot{\phi}} (t,L)  = \frac{1}{2}\braket{\{\dot{\phi}(t,u,\varphi), \dot{\phi}(t,u^{\prime},\varphi^{\prime})\}}_c.
    \label{eq:PhiDotPhiDotCorrelatorDefinition}
\end{equation}
The relation given in \eqref{eq:DensityDensityCorrelatorLeadingOrder} is a result of the normalization chosen in \eqref{eq:RescaledDensityContrast}. 

Moreover, as a consequence of the spatial homogeneity of FLRW universes, two-point correlation functions do not depend separately on the two spatial positions $(u, \varphi)$ and $(u^\prime, \varphi^\prime)$, but only on the comoving distance $L$ between them,
\begin{widetext}
    \begin{equation}
        L = \begin{cases}
        \frac{1}{\sqrt{\abs{\kappa}}}\cos^{-1}\left(\cos\theta\cos\theta^{\prime} + \sin\theta \sin\theta^{\prime} \cos(\varphi - \varphi^{\prime})\right) &\text{for} \hspace{0.1cm} \kappa > 0, \\
        \left[u^2+u^{\prime 2}-2uu^{\prime}\cos(\varphi-\varphi^{\prime})\right]^{1/2} &\text{for} \hspace{0.1cm} \kappa = 0, \\
        \frac{1}{\sqrt{\abs{\kappa}}} \cosh^{-1}\left(\cosh \sigma \cosh \sigma^{\prime} - \sinh \sigma \sinh \sigma^{\prime} \cos(\varphi-\varphi^{\prime})\right) &\text{for} \hspace{0.1cm} \kappa < 0,
        \end{cases}
        \label{eq:ComovingDistance}
    \end{equation}
\end{widetext}
and through \eqref{eq:DensityDensityCorrelatorLeadingOrder} the observable defined in \eqref{eq:RescaledDensityContrast} acquires the symmetries of the FLRW universe, 
\begin{equation}
    \mathcal{G}_{n n} (t; u, \varphi, u', \varphi') \equiv \mathcal{G}_{nn} (t,L).
\end{equation}
We proceed with the evaluation of this correlation function through \eqref{eq:PhiDotPhiDotCorrelatorDefinition} 
within the FLRW universe paradigm using the Bogoliubov transformations introduced in Sec.\ \ref{subsec:BogoliubovTrafo}. This leads to
\begin{equation}
    \mathcal{G}_{n n} (t, L) = \frac{\hbar a_\text{f}}{\bar{n}_0 m} \int_k \, \mathcal{F}(k, L) \sqrt{-h(k)} S_k (t),
    \label{eq:DensityDensityCorrelatorSpectrum}   
\end{equation}
where we introduced the spectrum of fluctuations
\begin{equation}
    S_k(t) = \frac{1}{2} + N_k + \Delta N_k (t),
    \label{eq:defSkt}
\end{equation}
as the momentum space representation of the rescaled density contrast two-point correlation function.

Therein, we have the expected occupation number of phonon excitations per mode
\begin{equation}
   N_{k} = \bra{\Omega}\hat{b}^{\dagger}_{km} \,  \hat{b}_{km} \ket{\Omega} = |\beta_k|^2,
   \label{eq:Spectrum}
\end{equation}
and the time-dependent contribution
\begin{equation}
    \Delta N_k(t) = \text{Re}\left[ c_k e^{2i\omega_kt}\right],
    \label{eq:defDeltaN}
\end{equation}
wherein
\begin{equation}
    c_k = - (-1)^m \bra{\Omega}\hat{b}^{\dagger}_{km} \, \hat{b}^{\dagger}_{k,-m}\ket{\Omega} = \alpha_k\beta_k.
    \label{eq:OffDiag}
\end{equation}
Furthermore, we used the abbreviation
\begin{equation}
    \int_{k}  = \begin{cases}
    \sum_{l=0}^{\infty} \abs{\kappa}\frac{l+1/2}{2\pi} &\text{for} \hspace{0.2cm} \kappa > 0, \\
    \int \frac{\text{d}k}{2\pi} \, k  &\text{for} \hspace{0.2cm} \kappa = 0,\\
    \int \frac{\text{d}l}{2\pi} \,\abs{\kappa} l \tanh(\pi l) &\text{for} \hspace{0.2cm} \kappa < 0,
    \end{cases}
\end{equation}
and the integration kernels
\begin{equation}
    \hspace{-0.15cm}\mathcal{F}(k, L) = \begin{cases}
    P_l\left(\cos{\left(L \sqrt{\abs{\kappa}}\right)}\right) &\text{for} \hspace{0.2cm} \kappa > 0, \\
     J_0\left(kL\right) &\text{for} \hspace{0.2cm} \kappa = 0, \\
    P_{il - 1/2}\left(\cosh{\left(L \sqrt{\abs{\kappa}}\right)}\right) &\text{for} \hspace{0.2cm} \kappa < 0,
    \label{eq:Kernels}
    \end{cases}
\end{equation}
for the different spatial curvatures specified by $\kappa$.

We can rewrite the spectrum of fluctuations \eqref{eq:defSkt} using \eqref{eq:defDeltaN} and Euler's formula
\begin{equation}
    S_k (t) = \frac{1}{2} + N_k + \abs{c_k} \cos \left(2 \omega_k t + \theta_k \right),
    \label{eq:SktWithPhase}
\end{equation}
where 
\begin{equation}
    \theta_k = \text{Arg} (c_k)
    \label{eq:PhaseDef}
\end{equation}
denotes the phase corresponding to the momentum mode $k$.

It is useful to note at this point that $N_k$ and $\Delta N_k$ defined in \eqref{eq:Spectrum} and \eqref{eq:defDeltaN} with the relations \eqref{eq:BogoliubovInTermsOfWroskian} are invariant under phase transformations on the mode functions $v_k(t) \to e^{i\lambda_k}v_k(t) $, $u_k(t) \to e^{i\mu_k} u_k(t)$, if one also takes into account $e^{2i\omega t}\to e^{-2i\mu_k}e^{2i\omega t}$ in Eq.\ \eqref{eq:defDeltaN}. These phases should not be observable.

Also, following \cite{Robertson2017a,Robertson2017b} one can show that entanglement between modes with opposite wave numbers can be witnessed in the two-mode squeezed state of relevance whenever $\Delta N_k > N_k$. The latter was recently observed experimentally in \cite{Chen2021} within a homogeneous two-dimensional Bose-Einstein condensate after quenching the time-dependent coupling $\lambda (t)$ to negative values and back. We leave a similar analysis within the FLRW universe paradigm for future work.

Moreover, it is important to note that a two-point correlation function of fields as defined in \eqref{eq:PhiDotPhiDotCorrelatorDefinition} shows an ultraviolet divergence \cite{Mukhanov2007}. Consequently, the rescaled density contrast correlation function \eqref{eq:DensityDensityCorrelatorSpectrum} has to be regularized. In the context of a Bose-Einstein condensate, a regularization arises naturally as the readout of the density contrast is limited by the precision of the measurement apparatus. Moreover, the acoustic approximation that we use here is a low momentum effective description that looses validity in the ultraviolet regime.
 
To solve this problem, which is well known in cosmology, one may work with smeared-out fields \footnote{Note that the case of $d=1+1$ spacetime dimensions is somewhat special from a cosmological point of view because the very large conformal group allows there often a map to flat space, hence spatial curvature is excluded.}
\begin{equation}
    \Phi(t,\vec{r})=\int \text{d}^2 r^\prime W (\vec{r}-\vec{r}') \phi(t,\vec{r}'), 
\end{equation}
with a window function $W (\vec{r}-\vec{r}')$. The latter can be normalized according to
\begin{equation}
    \int d^2 r^\prime W(\vec r^\prime) = 1.
\end{equation}
In momentum space, this window function plays the role of an ultraviolet regulator. We end up with a regularized expression for the rescaled density contrast correlation function
\begin{equation}
    \begin{split}
        G_{nn} (t,L) = \frac{\hbar a_{\text{f}}}{\bar{n}_0 m} \int_k \mathcal{F}(k,L) \sqrt{-h(k)} S_k (t) 
        \tilde{f}_G(k),
    \end{split}
    \label{eq:DensityDensityCorrelatorConvoluted}
\end{equation}
where $\tilde{f}_G(k) = \tilde{W}^* (k)\tilde{W} (k)$ corresponds to the absolute square of the Fourier transformed window function.

In the following we work with a window function of Gaussian form in position space (as a function of the comoving distance), such that in the absence of spatial curvature
\begin{equation}
  \tilde f_G(k) = \tilde W^*(k)\tilde W(k)  = e^{-w^2 k^2}.
   \label{eq:GaussianRegulator}
\end{equation}

\subsection{Stimulated particle production}
\label{subsec:StimulatedProduction}

An initial state at time $t_\text{i}$ with nonvanishing occupation number, such as a thermal state, would lead to stimulated particle production \cite{Calzetta2008}. This leads to a generalization of the expressions derived above. 

Assuming that the expected occupation number of quasiparticles originally present in the mode $k$ is given by
\begin{equation}
    \begin{split}
        N^{\text{in}}_k&=\braket{\hat{a}_{km}^{\dagger}\hat{a}_{km}},
    \end{split}
    \label{eq:Occupation}
\end{equation}
the total expected number of quasiparticles after expansion would be given by
\begin{equation}
    N_k = N_k^{\text{in}} + \abs{\beta_k}^2 \left(1+2 N_k^{\text{in}}\right),
\end{equation}
where the last term corresponds to the stimulated production. 
Similarly, $\Delta N_k(t)$ is generalized to
\begin{equation}
    c_k = \alpha_k\beta_k  \left(1 + 2 N_k^{\text{in}}\right).
\end{equation}
As an example we will consider a thermal state, characterized by an initial occupation number of the form 
\begin{equation}
    N^{\text{in}}_k (T) = \frac{1}{e^{\hbar \omega^{\text{I}}_k/(k_B T)}-1},
    \label{eq:Thermal}
\end{equation}
where $T$ denotes the temperature. 

In the following, to set a temperature scale, we use the critical temperature $T_c$ of an ideal gas in an anisotropic trap. In particular, we consider a ratio between longitudinal and radial trapping frequencies that elicit the emergence of a 2D condensate \cite{Dalfovo1999}. This critical temperature is given by 
\begin{equation}
    T_c = \frac{\hbar \omega }{k_B}\left(\frac{N}{\zeta(2)} \right)^{1/2},
    \label{eq:CriticalTemperature2D}
\end{equation}
where $N$ is the total number of atoms.

\begin{figure*}[t!]
\centering
\includegraphics[width=0.95\textwidth]{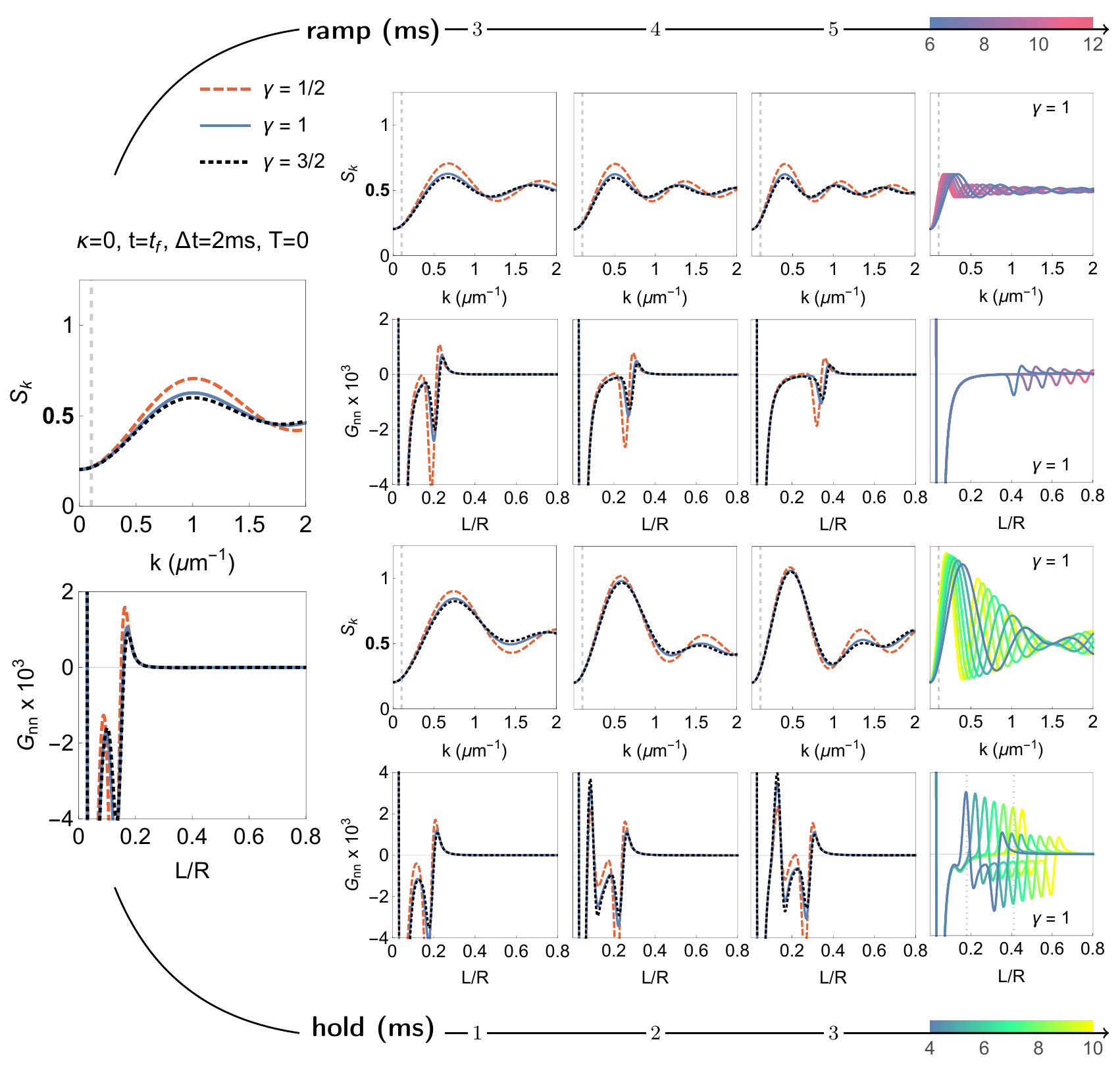}
\caption{Spectrum of fluctuations $S_k$ as a function of the radial wave number $k$ together with its corresponding rescaled density contrast correlation function $G_{nn}(L)$ (cf. Eq.\ \eqref{eq:DensityDensityCorrelatorConvoluted}) as a function of the comoving distance $L$ measured in units of the parameter $R$. Both are shown for various types of polynomial expansions (decelerating $\gamma=1/2$, uniform $\gamma=1$, and accelerating $\gamma=3/2$). Left column: Spectrum and correlation function right at the end of expansion. The dependence on different expansion rates $\Delta t$ is depicted above, and evolution after expansion is shown below. At the level of the spectra, one can appreciate that a slower expansion moves power to smaller wave numbers, whereas for faster expansions, more modes with higher momenta get excited. The hold time dependence shows how this power is evolving with time, in favor of the lower momentum modes. At the level of the correlation functions one can see that both, faster expansion rates $\Delta t$ and decelerated expansions ($\gamma < 1$) lead to stronger short-range anticorrelations. The propagation of the spatial correlations in time after the expansion has ceased is governed by the speed of sound in the condensate. In the lower row we show this: the correlation first builds up to a maximum, in this case reached at $3$ ms holding time, and then it travels through the condensate at twice the speed of sound. In the lower right panel we highlight through dotted lines the distance traversed after $5$ms by moving at twice the speed of sound. In all the momentum space plots, a gray vertical dashed line indicates the low $k$ limit at inverse condensate size. The position space results are obtained after regularization with a Gaussian window function of width $w= 0.5 \,\mu$m.}
\label{fig:RampAndHoldDependence}
\end{figure*}

\section{Effects of Expansion}
\label{sec:Effects}

Let us now focus on particle production as the main trait of an expanding spacetime, we do this within the developed formalism for a set of cosmological situations. In particular, we consider the spectrum of fluctuations together with the rescaled density contrast correlation function and discuss the outcome of various experimental scenarios. If not stated differently, the experimental values are taken from Appendix \ref{app:Values}.

\subsection{Polynomial scale factors with various expansion rates and holding times}

We base our analysis on polynomial scale factors of powers $\gamma$ according to eq.\ \eqref{eq:ScaleFactorPolynomial} which comprise accelerating ($\gamma = 3/2$), uniform ($\gamma = 1$), and decelerating ($\gamma = 1/2$) expansions, and look into different expansion rates $\Delta t$ and
hold times after the expansion has ceased. 

We study in Fig.\ \ref{fig:RampAndHoldDependence} the effect of increasing the expansion duration (at fixed ratio $a_\text{f}/a_\text{i}=\sqrt{6}$), and obtain an analogous prediction to a cosmological situation: for slower expansion the characteristic features of the power spectrum appear at smaller wave numbers. This can be read out from the change in the shape of the spectrum in the first row of Fig.\ \ref{fig:RampAndHoldDependence}. It can also be seen that a decelerating expansion leads to slightly higher contrast in the spectrum compared to accelerating or uniform expansion. Furthermore, at large momenta, the spectrum converges to the ground state or ``vacuum'' expression. In the second row of  Fig.\ \ref{fig:RampAndHoldDependence} we show the manifestation of these features in position space, through the correlation function $G_{nn} (t,L)$ given in \eqref{eq:DensityDensityCorrelatorConvoluted}. On top of a strong (diverging at $L=0$) anticorrelation coming from features of the ground state or  ``vacuum'', we see an anticorrelation-correlation pair at finite distances. The magnitude of this pair of correlations decreases with slower expansion rates, and is highest for a decelerating scenario.

\begin{figure*}[t!]
\centering
\includegraphics[width=0.95\textwidth]{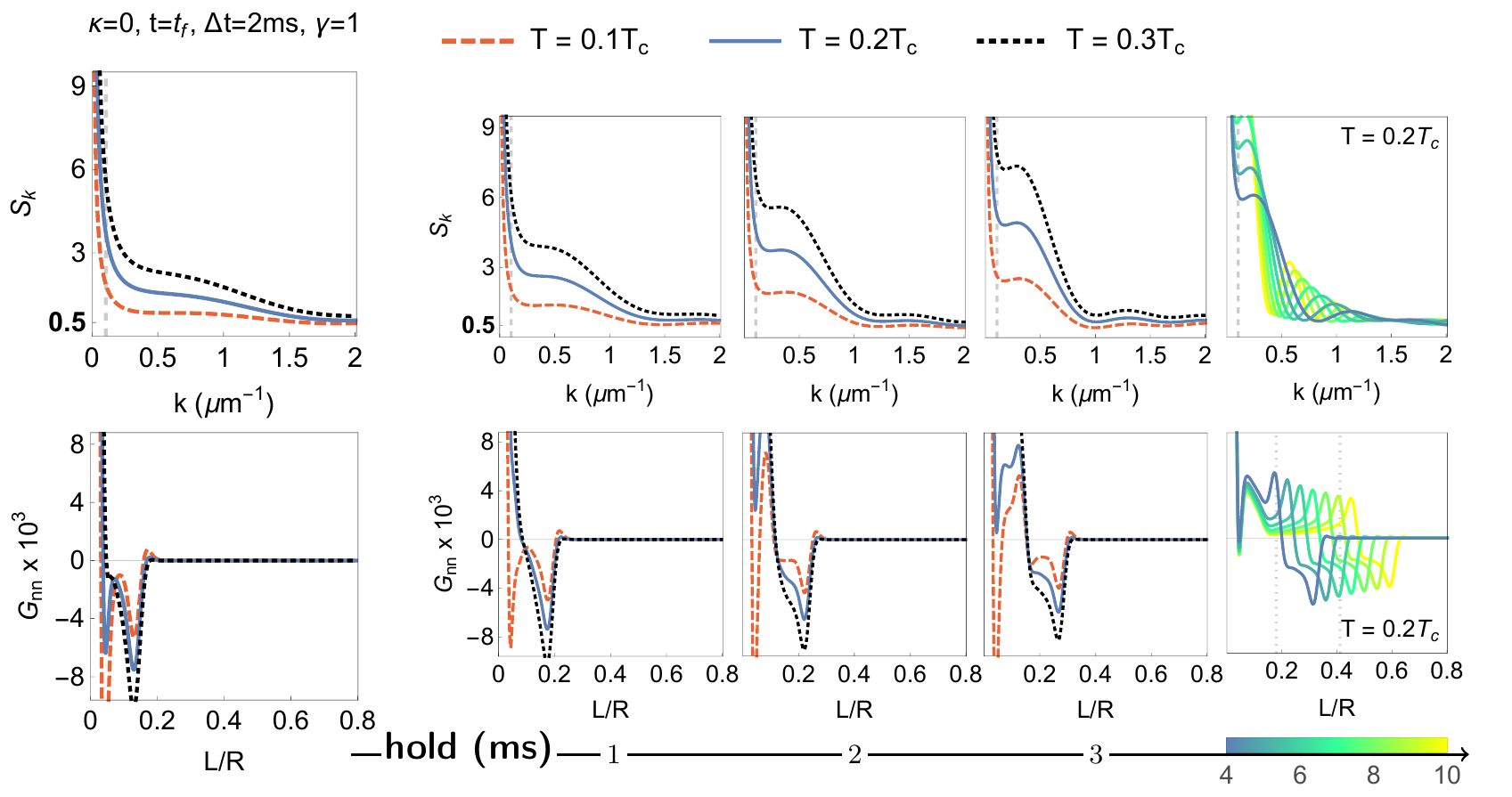}
\caption{Effect of initial temperature, leading to stimulated particle production: it can be seen that in this case the number of excited soft momenta is large, in contrast to particle production from the vacuum; note also that the spectrum $S_k$ diverges for $k \to 0$ as a consequence of the bosonic nature of gapless phonons at vanishing spatial curvature. After the expansion has ceased, the power apparently moves towards lower momentum modes, and the spectrum distribution oscillates around $S_k=(1/2 + \abs{\beta_k}^2)(1+2N_k^{\text{in}})$. Regarding position space, a finite initial occupation enhances anticorrelations and suppresses correlations. The short range correlation that builds up after the expansion has ceased is also enhanced, and the correlations propagate again at twice the speed of sound, indicated through dotted lines in the bottom right panel. The two point rescaled density contrast converges at long times to a thermal state, plus a finite contribution from the exited modes, determined by $\abs{\beta_k}^2$. In all the momentum space plots, a gray vertical dashed line indicates the low $k$ limit at inverse condensate size. The position space results given here correspond to a Gaussian convolution of  $w= 0.5 \,\mu$m standard deviation.}
\label{fig:ThermalState}
\end{figure*}

The evolution after expansion is given in the two lower rows of Fig.\ \ref{fig:RampAndHoldDependence}. After a simulated expansion, done by a $2$ ms ramp, has ceased, the spectrum shifts to lower momenta, and oscillates in time for each $k$ mode around $N_k$, with a period corresponding to the frequency of each mode. A node in the spectrum appears for a particular value of $k$ which is not excited in the process of particle production ($\beta_k = 0$). This precise feature is present only for the case of uniform expansion, and will be discussed further in Sec.\ \ref{subsec:MomentumModes}. 

Complementary, the position space evolution exhibits a decreasing magnitude of correlation-anticorrelation pair through time, along with a propagation to larger distances. In the short range we also see a correlation build up as a reaction to the expansion dynamics. The group of correlations propagates at twice the speed of sound (Eq.\ \eqref{eq:SpeedOfSound} at the center of the trap), as phonons travel away from each other. The chosen width for convolution (here $w= 0.5 \,\mu$m) influences the shape of  $G_{nn} (t,L)$, but not the position of the peaks, with the only exception being the vacuum anticorrelation, which goes to $L=0$ in the limit of vanishing width. Robust features with respect to changes of the width are discussed in Sec.\ \ref{subsec:RobustFeatures}.

\subsection{Initial thermal state}
\label{sec:ThermalInitialState}

Together with an initial thermal state comes about the phenomenon of stimulated particle production described in Sec.\ \ref{subsec:StimulatedProduction}. Given the divergence of the statistical distribution \eqref{eq:Thermal} at low momenta, there is a large occupation of soft modes also after the expansion, in contrast with the outcome for vanishing initial temperature. Since stimulated particle production is present for any state above the ground state with $T=0$, the study of this phenomenon is crucial when comparing to realistic experimental situations. On the other hand, the question arises whether the initial state in a concrete experiment is actually thermal. In any case, the phenomenon of particle production can help to investigate the properties of the initial state in an experimental setting. 

In Fig.\ \ref{fig:ThermalState} we investigate stimulated particle production, for three different initial temperatures in fractions of $T_c$, at the level of both, the spectrum and the rescaled density contrast correlation function as a function of hold time.

\begin{figure}[t!]
\centering
\includegraphics[width=0.45\textwidth]{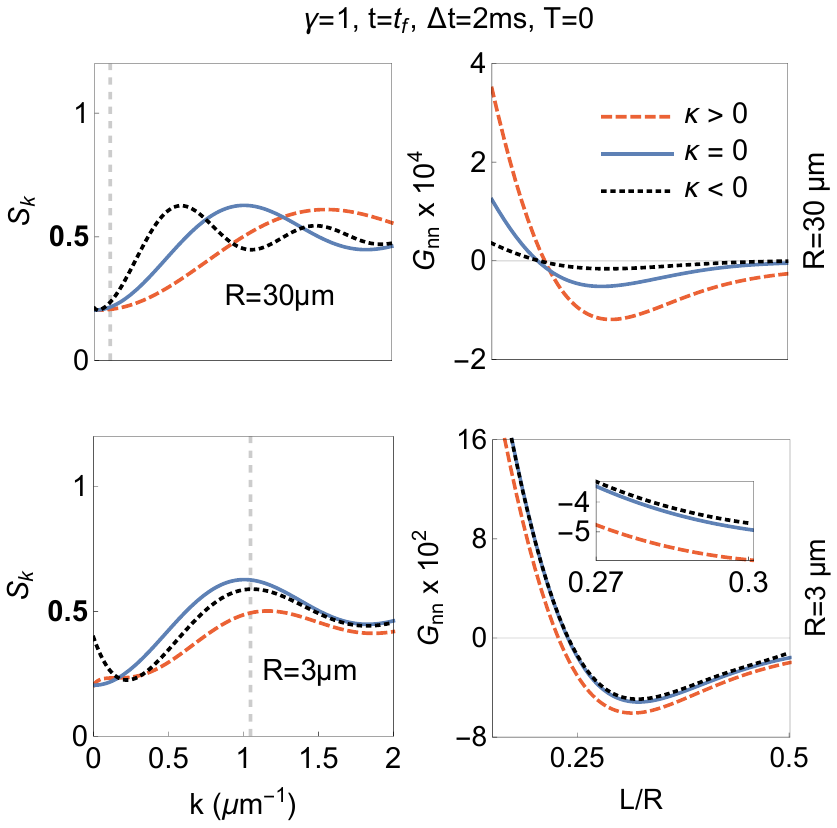}
\caption{Effect of different trapping potentials (inverted harmonic, box, harmonic), which convey different spatial curvatures (positive, flat, negative; respectively). On the upper row we show the outcome for a fixed atom number, so that the density at the center of the trap depends on the shape of the trap (c.f. \eqref{eq:AtomNumber} and \eqref{eq:TrapShape}); this renders visible differences for both the spectrum $S_k$ and the rescaled density contrast $G_{nn}$ (obtained with a Gaussian convolution of width $w=4 \, \mu$m). However, if the density at the center is taken to be the same for all trap shapes, the spatial curvature $\kappa$ influences the shape of the spectrum $S_k$ only at low momentum modes and provided that $R$ is sufficiently small, as depicted in the lower row. The differences between rescaled density contrast correlations for the closed, flat, and open universe are barely visible even when $R=3\, \mu$m. Additionally, for the chosen convolution width ($w=0.4 \, \mu$m) the vacuum sector dominates in position space. An inset into the correlation function shows that the results for each curvature fall one above the other: this is a width independent feature.}
\label{fig:Kappa}
\end{figure}

\subsection{Spatial curvature}

We showed in Sec.\ \ref{subsec:MetricToFLRW} that different types of trapping potentials induce acoustic spacetimes with different emergent spatial curvatures. In the formalism employed in Sec.\ \ref{subsec:ModeFunctions}, the effects of spatial curvature are carried into the shape of Laplace-Beltrami's operator, its eigenfunctions, and its eigenvalues. Regarding time evolution in momentum space (cf. Eq.\ \eqref{eq:ModeEquation}), spatial curvature enters in fact only through the eigenvalue spectrum. Here, the features of spatial curvature are equivalent to posing different boundary conditions on the eigenvalue equation, and do not go further than that. A further dependence on spatial curvature arises in the integral transform from momentum to position space (cf. Eq.\ \eqref{eq:Kernels}).

As expected, the effect of curvature on the spectrum of fluctuations is often negligible, but can be tuned to a higher impact when decreasing the condensate radius. Something similar happens to the rescaled density contrast, where differences are unimportant, even at small radii; this is shown in Fig.\ \ref{fig:Kappa}. In the presence of an initial thermal state this situation could change, given that the Bose-Einstein distribution \eqref{eq:Thermal} differs for different dispersion relations. In particular for negative curvature $N_k^{\text{in}}$ is bounded at $k=0$, as a consequence of an acquired gap in the dispersion relation. This was investigated and no particular differences were found at different curvatures.

\begin{figure}[t!]
\includegraphics[width=0.45\textwidth]{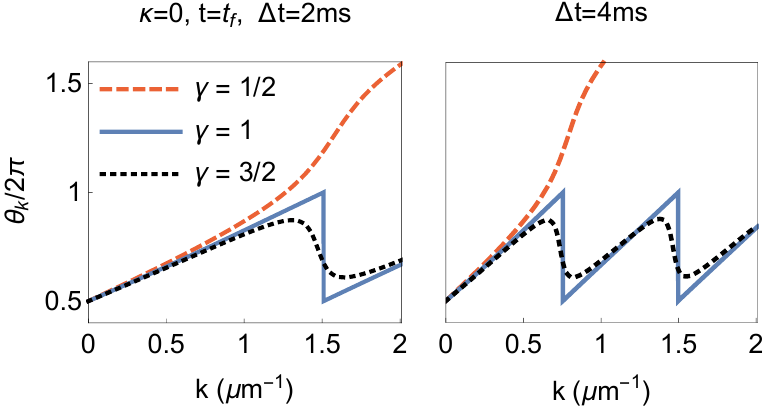}
\caption{Phases $\theta_k$ (cf. \eqref{eq:PhaseDef}) of each $k$-mode after expansion for three different scale factors. We see an important qualitative difference in behavior depending on the choice of $\gamma$: for $\gamma = 1$ and $\gamma = 3/2$ the phase remains in a range of $(0,\pi)$, in particular, in the $\gamma = 1$ case the phase is a linear function of $k$. Also for this case there is a $\pi$ jump which appears if and only if $N_k = 0$, that is when a particular mode is not excited as a result of expansion. The other two scale factors don't exhibit this feature. In contrast, for $\gamma = 1/2$ the phase increases continuously in the complete range $(0,2\pi)$. In all cases, a slower expansion rate allows for a greater phase growth.}
\label{fig:Phases}
\end{figure}

\subsection{Time evolution of momentum modes and robust features in momentum space}
\label{subsec:MomentumModes}

\begin{figure*}[t!]
\includegraphics[width=0.95\textwidth]{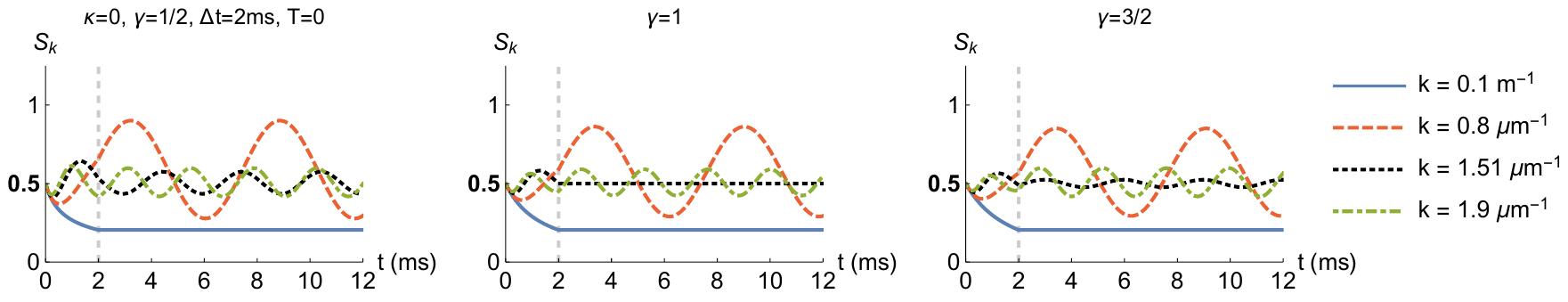}
    \caption{Time evolution of the spectrum $S_k$ for four radial wave numbers $k$ from $t=0$ (setting $t_i = 0$) up to $t = t_{\text{f}} + 10$ ms. Emphasized through a vertical dashed line is the moment the expansion ceases, i.e. $t = t_{\text{f}}$. The low momentum modes are suppressed by the expansion, and they almost become static when the expansion has ceased. For higher momentum modes the amplitude at the time of expansion is greater in the case of a decelerating scenario. Depending on the duration of expansion there are always certain modes (in this case $k=1.51 \,\mu\text{m}^{-1}$) which remain in its vacuum state after expansion for the $\gamma=1$ situation. These same modes exhibit the greatest difference in amplitude between the $\gamma =1/2$ and $\gamma =3/2$ situations.}
    \label{fig:SpectraEvolution}
\end{figure*}

As we see overall in Fig.\ \ref{fig:RampAndHoldDependence}, the qualitative differences related to different exponents $\gamma$ in the scale factor could be difficult to appreciate experimentally, when looking into the complete spectrum and rescaled density contrast. Nevertheless, one can look into details of the spectrum, and through them, validate the particle production nature of the experimental outcomes, and its dependence on different expansion histories.

We consider first the time evolution of the spectra for certain modes in regions II and III (cf. Fig.\ \ref{fig:SpectraEvolution}) that is, during and after the dynamic change of the scale factor; there we observe that, independently of the polynomial power of the scale factor, the phononic modes corresponding to small wave numbers are suppressed by the expansion, which is consistent with Fig.\ \ref{fig:RampAndHoldDependence}. Moreover, the time evolution of the each momentum mode shows a slight dependence on the expansion history, which is most evident for a particular mode, at $k=1.51 \, \mu\text{m}^{-1}$, that remains in its vacuum value after uniform expansion ($\gamma = 1$), given that this characteristic is not present for any mode in nonuniform expansions. 

This precise feature is also explicit in the phase $\theta_k$ that each mode acquires after expansion, defined in \eqref{eq:PhaseDef}, which we consider next in Fig.\ \ref{fig:Phases}. 
There, we see that the phases of each wave number $k$ strongly depend on whether there is a decelerated, uniform, or accelerated expansion. In the case of uniform expansion there are phase jumps appearing at each mode where $\beta_k$ turns out to be zero. To emphasize, due to the shape of expansion, $\beta_k$ is never zero for $\gamma \neq 1$. It is worthwhile to note that phases as a function of $k$ also give an insight into the expansion duration $\Delta t$.

\begin{figure}[t!]
\centering
\includegraphics[width=0.42\textwidth]{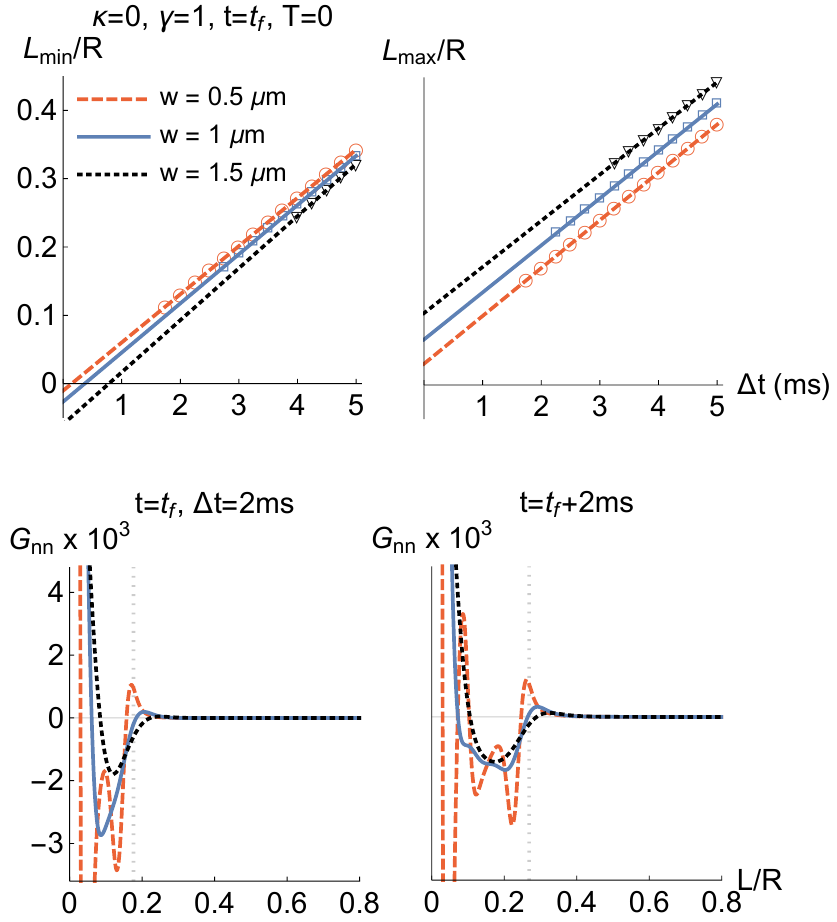}
\caption{Upper row: positions of the second minimum $L_{\text{min}}/R$ (left) and first maximum $L_{\text{max}}/R$ (right) of the rescaled density contrast correlation function $G_{n n} (t,L)$ as a function of the expansion duration $\Delta t$ for various widths $w$. The linear fits exhibit slopes robust against $w$ and almost fall on top of each other in the former case. Lower row: $G_{n n} (t,L)$ for various widths $w$ at the time the expansion has ceased (left panel), and two milliseconds after (right panel). Most sensible to choice of $w$ are the position and magnitude of the vacuum anticorrelation. The position of the group of correlations coming from particle production is relatively robust against $w$, but the peaks tend to disappear as the width grows closer to the healing length $\xi \approx 1.6 \,\mu$m. However, the group of correlations moves with twice the speed of sound, independent of the width $w$ (indicated by a gray dotted line).} 
\label{fig:RobustPosition}
\end{figure}

\subsection{Window function dependence and robust features in position space}
\label{subsec:RobustFeatures}

Given that any computation of the rescaled density contrast correlation function requires an ultraviolet regulator in the form of a window or test function, we wish to find features of the latter which are robust against variations in the standard deviation $w$ of the Gaussian family of window functions we have chosen in Eq.\ \eqref{eq:GaussianRegulator}. This is not only an interesting task by itself, but also paves the ground for a quantitative comparison to experiments.

To that end, we study the positions of the second minimum --the first minimum is just the vacuum contribution-- and the first maximum of the correlation function. More precisely, we investigate the aforementioned positions as a function of expansion duration $\Delta t$ for different widths $w$, for the particular case of $\gamma=1/2$. The results are shown in Fig.\ \ref{fig:RobustPosition}: we find that the influence of the width $w$ is negligible with regards to the slope of the curves, rendering position vs. expansion duration a robust observable. Also in Fig.\ \ref{fig:RobustPosition}, the rescaled density contrast correlation function $G_{n n} (t,L)$ is shown for different widths $w$. The resolution $w$ determines the short-length $L < 0.1 R$ behavior of the two-point correlator indicating the need for robust features.

Moreover, let us report that we have also investigated the amplitudes of the maximum and the minimum and their ratio, but did not find a similar form of robustness.

\section{Conclusion and Outlook}
\label{sec:Outlook}
In summary, we have derived a correspondence between phonons in a $2+1$ dimensional Bose-Einstein condensate in various radially symmetric trapping potentials and massless scalar particles in spatially curved FLRW universes. As opposed to common literature, this correspondence was established starting from a nonrelativistic action and describing phononic excitations in terms of the real and imaginary parts of the fluctuations on top of the mean field.

Furthermore, we investigated the phenomenon of particle production in momentum and in position space for various experimentally accessible scenarios. We showed that a suitably rescaled density contrast correlation function is to leading order proportional to a correlator within the FLRW universe paradigm. As a consequence of rescaling, spatial homogeneity and isotropy carried over to the density contrast correlation function. Looking into experimental feasibility, we have shown that the phases of momentum modes and the positions of maxima and minima in the correlation function serve as robust observables and are distinguishable for different dynamics in the scale factor.

In future theoretical work it would be interesting to extend our approach from $d=2+1$ to $d=3+1$ dimensions, which should be straightforward. Also, one may want to study more general excitation fields, involving spin degrees of freedom, or modes with gaped excitation spectrum. Also a generalization to the full Bogoliubov spectrum beyond the acoustic approximation can be of interest for some questions. 

Another interesting direction could be to investigate other expansions specified by the scale parameter, allowing for a simulation of various epochs of expanding or contracting universes. In particular, one may study the de Sitter universe in the context of inflation or even a cyclic universe, probably exhibiting additional features such as parametric resonance.

Moreover, one may investigate particle production from a quantum information theoretic perspective. More precisely, entanglement between modes of opposite momenta may be analyzed for spatially curved universes, extending the work of \cite{Robertson2017a,Robertson2017b}.

Most importantly, it is of great interest to apply our methods to concrete ultra cold atoms experiments. In particular, as the rescaled density contrast correlation function serves as a typical observable, our work paves the ground to an extensive experimental investigation of particle production. We report on experimental confirmation of our proposal in \cite{ExperimentPaper2022}.

\section*{Acknowledgements}
The authors would like to thank Simon Brunner and Finn Schmutte for useful discussions. This work is supported by the Deutsche Forschungsgemeinschaft (DFG, German Research Foundation) under Germany's Excellence Strategy EXC 2181/1 - 390900948 (the Heidelberg STRUCTURES Excellence Cluster) and under SFB 1225 ISOQUANT - 273811115, the ERC Advanced Grant Horizon 2020 EntangleGen (Project-ID 694561) as well as FL 736/3-1. NSK is supported by the Deutscher Akademischer Austauschdienst (DAAD, German Academic Exchange Service) under the Länderbezogenes Kooperationsprogramm mit Mexiko: CONACYT Promotion, 2018 (57437340). APL is supported by the MIU (Spain) fellowship FPU20/05603 and the MICINN
(Spain) project PID2019-107394GB-I00 (AEI/FEDER,
UE). NL acknowledges support by the Studienstiftung des Deutschen Volkes.\\

%\vspace{1mm}
%\noindent
%\textbf{Author contributions}: S.F., T.H., N.S-K, A.P-L., M.T-S. built the theoretical framework. Experimental feasibility was developed in discussion among all authors. 
%All authors contributed to the discussion of the results and the writing of the manuscript.

%M.H., E.K., N.L., M.K.O., M.S., H.S., C.V.

%%%%%%%% Appendix %%%%%%%%

\appendix

\section{Full Bogoliubov dispersion relation}
\label{app:BogoliubovDispersionRelation}
In order to asses the validity of the acoustic approximation, it is also interesting to perform calculations in full Bogoliubov theory for excitations in weakly interacting Bose-Einstein gases. This is applicable in particular for homogeneous Bose-Einstein condensates, and for static situations such as quantum fluctuations around the ground state, or in thermal equilibrium. An extension of this formalism to time-dependent situations and more general trapping potentials is possible, but beyond our scope in the present paper.
  
The Bogoliubov dispersion relation for quasiparticles reads
\begin{equation}
    \omega_k = \frac{\hbar}{2 m} \sqrt{k^2 (k^2 + 2/\xi^2)}.
    \label{eq:FullBogoliubovDispersionRelation}
\end{equation}
This features a transition at the healing length
\begin{align}
    \xi=\frac{\hbar}{\sqrt{2m \lambda n_0}},
\end{align}
such that for small wave numbers $k \ll 1/\xi$ the dispersion relation is linear, $\omega_k = c k$, with the speed of sound $c$ as defined in \eqref{eq:SpeedOfSound}, while for large wave numbers $k \gg 1/\xi$ it becomes quadratic, $\omega_k \to \hbar k^2 / (2m) + \lambda n_0 / \hbar$. 

\begin{figure*}[t!]
    \centering
    \includegraphics[width=0.95\textwidth]{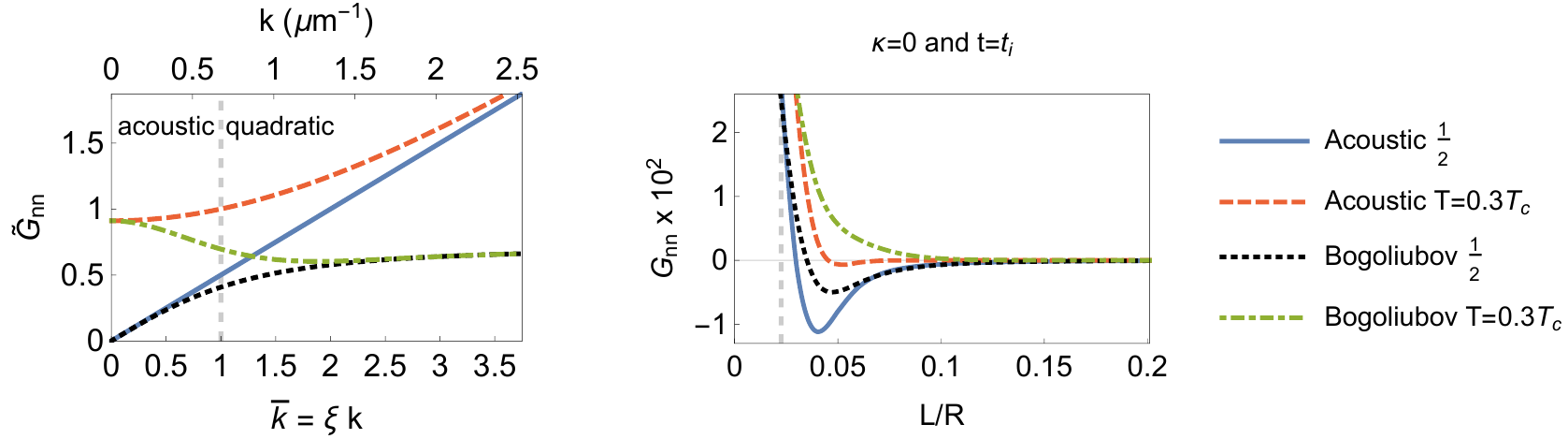}
    \caption{The influence of the full Bogoliubov dispersion relation \eqref{eq:FullBogoliubovDispersionRelation} and an initial temperature $T = 0.3 T_c$ is discussed at initial time $t=t_{\text{i}}$ for the rescaled density contrast correlation function in momentum space (left panel) and in position space (right panel). For small momenta $k \ll 1/\xi$ (indicated by the dashed vertical line), the Bogoliubov dispersion approaches the acoustic result (dashed vs. dot-dashed and solid vs. dotted curves), while for large momenta $k \gg 1/\xi$ the temperature dependence vanishes due to the exponential fall-off of the Bose-Einstein distribution $N_k^{\text{in}} (T)$ for large $k$ (dashed vs. solid and dotted vs. dot-dashed curves). Moreover, the correlation function is clearly influenced by a nonvanishing temperature $T=0.3T_c$, while a dependence on whether one includes the full Bogoliubov dispersion is subleading due to the Gaussian cutoff in momentum space as $w \approx \xi$ at initial time.}
    \label{fig:SpectraAndCorrelatorsBogoliubov}
\end{figure*}

For an initial thermal state with temperature $T < T_c$ introduced in \eqref{eq:Thermal}, the spectrum of fluctuations is given by
\begin{equation}
    S_k (t_{\text{i}}) = \frac{1}{2} + N_k^{\text{in}} (T),
\end{equation}
while the rescaled density contrast correlation function evaluates to \cite{Sanchez2021}
\begin{equation}
G_{n n} (t_{\text{i}},L) = \frac{2}{\bar{n}_0} \int_k k J_0 (k L) \sqrt{\frac{k^2}{k^2+ 2/\xi^2}} S_k (t_{\text{i}}) \tilde{f}_G (k), 
\end{equation}
where the linear dispersion relation $\sqrt{-h (k)} = k$ has been replaced by the corresponding expression of the full Bogoliubov dispersion relation \eqref{eq:FullBogoliubovDispersionRelation}. Note that in the acoustic limit $k \ll 1/\xi$, we obtain Eq.\ \eqref{eq:DensityDensityCorrelatorConvoluted} with $a_{\text{f}}$ replaced by $a_{\text{i}}$. The integrand and its respective correlator are shown in Fig.\ \ref{fig:SpectraAndCorrelatorsBogoliubov} for the acoustic vacuum and the Bogoliubov vacuum with and without an initial temperature $T = 0.3 T_c$.\\

\section{Experimental setup}
\label{app:Values}
The plots shown in Sec.\ \ref{sec:Effects} are computed for the following experimental parameters, setting $t_{\text{i}}=0$, out of convenience. The condensate consists of $N= 15 \times 10^3 $ potassium atoms of mass $m=6.47008 \times 10^{-26}$ kg and extends up to a radius $R= 30 \times 10^{-6}$ m. The longitudinal trapping frequency is $1.75 \times 10^3 \, \text{s}^{-1}$, while the initial and final $s$-wave scattering lengths are $a_s (t_{\text{i}})=300 a_0$ and $a_s (t_{\text{f}})= 50 a_0$, respectively, where $a_0$ denotes the Bohr radius. The 2D critical temperature for an anisotropic trap is calculated for these values at initial time, and yields $T_c= 82.41 \times 10^{-9}$ K. Moreover, the regularization for the correlation functions is carried out with a Gaussian of inverse standard deviation $w = 5 \times 10^{-7}$ m.\\

\newpage

%%%%%%%% Bibliography %%%%%%%%
\bibliography{references.bib}

\end{document}